\PassOptionsToPackage{pdfpagelabels=false}{hyperref}
\makeatletter
    \newcommand{\dontusepackage}[2][]{%
        \@namedef{ver@#2.sty}{9999/12/31}
        \@namedef{opt@#2.sty}{#1}
    }
\makeatother
\dontusepackage{fixltx2e}

\documentclass[fleqn,usenatbib]{mnras}

\usepackage{ae,aecompl}
\usepackage{amsmath,amssymb}
\usepackage{bbm}
\usepackage{booktabs}
\usepackage{datetime}
\usepackage{etoolbox}
\usepackage[T1]{fontenc}
\usepackage{graphicx}
\usepackage{mathtools}
\usepackage{multirow}
\usepackage{newtxtext}
\usepackage[varg,varvw]{newtxmath}
\usepackage{physics}
\usepackage{relsize}
\usepackage[alsoload=astro]{siunitx}
\usepackage{xcolor}
\usepackage{xparse}

\definecolor{MNRASpurple}{HTML}{BC3B9C}

\DeclareMathAlphabet{\mathsf}{OT1}{\sfdefault}{m}{n}
\DeclareMathAlphabet{\mathit}{OT1}{\itdefault}{m}{n}
\DeclareMathAlphabet{\mathcal}{OMS}{cmsy}{m}{n}
\SetMathAlphabet{\mathsf}{bold}{OT1}{\sfdefault}{b}{n}

\DeclareSIUnit\h{\text{$h$}}
\hypersetup{
    pdfauthor={Mike (Shengbo) Wang},
    pdftitle={Cosmological inference from galaxy-clustering power spectrum: Gaussianization and covariance decomposition},
    pdfsubject={Cosmology},
    pdfkeywords={methods: data analysis -- methods: statistical -- large-scale structure of Universe}
}

\newdateformat{yearmonthdate}{%
    \THEYEAR\ \monthname[\THEMONTH] \THEDAY
    }
\yearmonthdate

\newcommand{\appref}[1]{{appendix~\ref*{#1}}}

\makeatletter
    \newcommand{\linktarget}[1]{\Hy@raisedlink{\hypertarget{#1}{}}}
\makeatother

\let\oldeqref\eqref
\makeatletter
\RenewDocumentCommand\eqref{s m}{%
    \IfBooleanTF#1%
    {\textup{\tagform@{\ref*{#2}}}}%
    {\oldeqref{#2}}%
}
\makeatother

\makeatletter
\patchcmd{\NAT@citex}
    {\@citea\NAT@hyper@{%
    \NAT@nmfmt{\NAT@nm}%
    \hyper@natlinkbreak{\NAT@aysep\NAT@spacechar}{\@citeb\@extra@b@citeb}%
    \NAT@date}}
    {\@citea\NAT@nmfmt{\NAT@nm}%
    \NAT@aysep\NAT@spacechar\NAT@hyper@{\NAT@date}}{}{}
\patchcmd{\NAT@citex}
    {\@citea\NAT@hyper@{%
        \NAT@nmfmt{\NAT@nm}%
        \hyper@natlinkbreak{\NAT@spacechar\NAT@@open\if*#1*\else#1\NAT@spacechar\fi}%
        {\@citeb\@extra@b@citeb}%
        \NAT@date}}
    {\@citea\NAT@nmfmt{\NAT@nm}%
        \NAT@spacechar\NAT@@open\if*#1*\else#1\NAT@spacechar\fi\NAT@hyper@{\NAT@date}}
    {}{}
\makeatother

\renewcommand{\jcap}{J. Cosmol. Astropart. Phys.}

\DeclareMathOperator*{\argmax}{arg\,max}
\DeclareMathOperator*{\argmin}{arg\,min}
\DeclareMathOperator*{\Conv}{\mathlarger{\mathlarger{\mathrel{\raisebox{-1pt}{$\ast$}}}}}
\DeclareMathOperator{\Diag}{\textrm{diag}}
\DeclareMathOperator{\Erfc}{\textrm{erfc}}
\DeclareMathOperator{\Expc}{\mathbbmss{E}}
\DeclareMathOperator{\Var}{\textrm{Var}}
\DeclareMathOperator{\Prob}{\mathbbmss{P}}
\DeclareMathOperator{\Prior}{\Pi}
\DeclareMathOperator{\Like}{\mathcal{L}}
\DeclareMathOperator{\Post}{\mathcal{P}}
\DeclareMathOperator{\Norm}{\textrm{N}}
\DeclareMathOperator{\Wishart}{\textrm{W}}
\DeclareMathOperator{\WishartInv}{\textrm{W}^{-1}}

\newcommand{\biga}[1]{\big\langle#1\big\rangle}
\newcommand{\bigb}[1]{\big\lbrace#1\big\rbrace}
\newcommand{\bigs}[1]{\big[#1\big]}
\newcommand{\Bigs}[1]{\Big[#1\Big]}
\newcommand{\bigp}[1]{\big(#1\big)}

\newcommand{\biggp}[1]{\bigg(#1\bigg)}
\newcommand{\KL}[2]{D_\textrm{KL}\kern0.2pt(#1\kern0.4pt\|\kern0.4pt #2)}
\newcommand{\given}[2]{\kern-0.25pt\left.#1\kern1.25pt\middle|\kern1.25pt#2\right.\kern-0.25pt}
\newcommand{\bgiven}[2]{\kern-0.25pt#1\kern1.25pt\big|\kern1.25pt#2\kern-0.25pt}

\newcommand{\e}{\kern0.25pt\mathrm{e}\kern0.25pt}
\newcommand{\im}{\kern0.25pt\mathrm{i}\kern0.25pt}
\newcommand{\R}{\mathbbmss{R}}
\newcommand{\conj}[1]{#1^{*}}
\newcommand{\trans}[1]{#1^{\intercal}}
\newcommand{\overbar}[1]{\mkern2mu\overline{\mkern-2mu#1\mkern-2mu}\mkern2mu}
\newcommand{\est}[1]{\widehat{#1}}
\newcommand{\mat}[1]{\mathbfss{{#1}}}

\newcommand{\iz}{iz}
\newcommand{\gal}{\textrm{g}}
\newcommand{\simu}{\textrm{s}}
\newcommand{\data}{\textrm{d}}
\newcommand{\fid}{\textrm{f}}
\newcommand{\true}{\textrm{true}}
\newcommand{\shot}{\textrm{shot}}
\newcommand{\NL}{\textrm{NL}}

\newcommand{\den}{\delta}
\newcommand{\nbar}{\bar{n}}
\newcommand{\Pb}{\widetilde{P}_0}
\newcommand{\Pd}{\widehat{P}_0}
\newcommand{\winG}{\mathcal{G}}
\newcommand{\winW}{\mathcal{W}}
\newcommand{\thetap}{\bmath{\theta}}

\newcommand{\vk}{\vb*{k}}
\newcommand{\vq}{\vb*{q}}
\newcommand{\vr}{\vb*{r}}
\newcommand{\vx}{\vb*{x}}

\newcommand{\vmu}{{\bmath{\mu}}}
\newcommand{\X}{\vb*{X}}
\newcommand{\Y}{\vb*{Y}}
\newcommand{\Z}{\vb*{Z}}
\newcommand{\wY}{\widetilde{Y}}
\newcommand{\tY}{\widetilde{\Y}}
\newcommand{\mB}{\mat{B}}
\newcommand{\mC}{\mat{C}}
\newcommand{\mSig}{\boldsymbol{\mathsf{\Sigma}}}
\newcommand{\mLamb}{\boldsymbol{\mathsf{\Lambda}}}

\title[Gaussian{\iz}ation and covariance decomposition]{%
    Cosmological inference from galaxy-clustering power spectrum: %
    Gaussian{\iz}ation and covariance decomposition%
}
\author[M. S. Wang et al.]{%
    \parbox{\linewidth}{%
        Mike (Shengbo) Wang%
        \textsuperscript{\,\,\href{http://orcid.org/0000-0002-2652-4043}{\includegraphics[width=2.5mm]{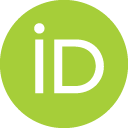}}},%
        \textsuperscript{\hyperlink{affil1}{1}}%
        \thanks{\hspace{-0.42em}Email: \href{mailto:mike.wang@port.ac.uk}{{mike.wang@port.ac.uk}}} %
        Will J. Percival,%
        \textsuperscript{\hyperlink{affil2}{2},\hyperlink{affil3}{3},\hyperlink{affil1}{1}} %
        Santiago Avila%
        \textsuperscript{\,\,\href{http://orcid.org/0000-0001-5043-3662}{\includegraphics[width=2.5mm]{ORCIDiD_icon128x128.png}}},%
        \textsuperscript{\hyperlink{affil1}{1}} %
        Robert Crittenden%
        \textsuperscript{\,\,\href{http://orcid.org/0000-0002-5743-1528}{\includegraphics[width=2.5mm]{ORCIDiD_icon128x128.png}}}%
        \textsuperscript{\hyperlink{affil1}{1}} \\
        and Davide Bianchi%
        \textsuperscript{\hyperlink{affil1}{4},\hyperlink{affil1}{1}}
    }
    \vspace{2.5mm} \\
    \!\!\linktarget{affil1}{\textsuperscript{\textup{1}}}Institute of Cosmology and Gravitation, University of Portsmouth, Burnaby Road, Portsmouth PO1~3FX, UK \\
    \!\!\linktarget{affil2}{\textsuperscript{\textup{2}}}Department of Physics and Astronomy, University of Waterloo, 200 University Avenue West, Waterloo, Ontario N2L~3G1, Canada \\
    \!\!\linktarget{affil3}{\textsuperscript{\textup{3}}}Perimeter Institute for Theoretical Physics, 31 Caroline Street North, Waterloo, Ontario N2L~2Y5, Canada \\
    \!\!\linktarget{affil4}{\textsuperscript{\textup{4}}}Institut de Ci\`{e}ncies del Cosmos, University of Barcelona, IEEC-UB, Mart\'{i} i Franqu\'{e}s, 1, E-08028 Barcelona, Spain
}
\date{Accepted 2019 March 18. Received 2019 March 15; in original form 2018 November 19}
\pubyear{2019}
\volume{486}
\pagerange{951--965}

\begin{document}

\maketitle

\begin{abstract}
    Likelihood fitting to two-point clustering statistics made from galaxy surveys usually assumes a multivariate normal distribution for the measurements, with justification based on the central limit theorem given the large number of overdensity modes. However, this assumption cannot hold on the largest scales where the number of modes is low. Whilst more accurate distributions have previously been developed in ideal{\iz}ed cases, we derive a procedure suitable for analysing measured monopole power spectra with window effects, stochastic shot noise and the dependence of the covariance matrix on the model being fitted all taken into account. A data transformation is proposed to give an approximately Gaussian likelihood, with a variance--correlation decomposition of the covariance matrix to account for its cosmological dependence. By comparing with the modified-$t$ likelihood derived under the usual normality assumption, we find in numerical tests that our new procedure gives more accurate constraints on the local non-Gaussianity parameter $f_\NL$, which is sensitive to the large-scale power. A simple data analysis pipeline is provided for straightforward application of this new approach in preparation for forthcoming large galaxy surveys such as DESI and \textit{Euclid}.
\end{abstract}


\defcitealias{Feldman:1993ky}{FKP}
\defcitealias{Sellentin:2015waz}{SH}

\section{\texorpdfstring{I\hspace{1.10pt}n\hspace{1.10pt}t\hspace{1.10pt}r\hspace{1.10pt}o\hspace{1.10pt}d\hspace{1.10pt}u\hspace{1.10pt}c\hspace{1.10pt}t\hspace{1.10pt}i\hspace{1.10pt}o\hspace{1.10pt}n}{Introduction}}
\label{sec:introduction}

The matter power spectrum, which measures the two-point correlation in Fourier space, is an important statistic for describing the large-scale structure of the Universe. It contains all of the information about a Gaussian random field, which describes cosmic density fluctuations on large scales where non-linearities are negligible. Galaxies are linearly biased tracers of the underlying matter distribution on large scales, and thus measurements of the comoving galaxy-clustering power spectrum can provide a wealth of information about fundamental cosmological parameters. Moreover, as the angular positions and redshifts of galaxies are observed, matching the expected comoving clustering offers a geometrical test of the Universe through the distance--redshift relationship, and measurements of redshift space distortions (RSD) provide a powerful probe of structure growth. Upcoming large-volume surveys, including the Dark Energy Spectroscopic Instrument\footnote{\hspace{-0.42em}\href{https://www.desi.lbl.gov/}{https://www.desi.lbl.gov/}.} \citep[DESI Collaboration;][]{Aghamousa:2016zmz} and \textit{Euclid}\footnote{\hspace{-0.42em}\href{https://www.euclid-ec.org/}{https://www.euclid-ec.org/}.} \citep[Euclid Consortium;][]{Laureijs:2011}, will be able to tightly constrain cosmological models with unprecedented precision, but the accuracy of these constraints relies on performing careful statistical analyses.

The multivariate normal distribution is ubiquitous in modelling cosmological observables thanks to the central limit theorem, and this normality assumption is commonly found in likelihood analyses of power spectrum measurements from galaxy surveys \citep[e.g.][]{Alam:2016hwk,Abbott:2017wau}. Given a theoretical model for the data, the key ingredient of a multivariate normal distribution is its covariance matrix, and in the past many efforts have been devoted to the accurate estimation of covariance matrices subject to limited computational resources.

Unbiased covariance matrix estimates are often made from a set of mock galaxy catalogues synthes{\iz}ed using algorithms ranging from fast but approximate perturbation-theory algorithms to slow yet detailed $N$-body simulations, or different combinations of those \citep[e.g.][]{Manera:2013xx,Kitaura:2015uqa,Avila:2017nyy}. An overview and comparison of those methods is provided in a series of papers by \cite{Lippich:2018wrx}, \cite{Blot:2018oxk}, and \cite{Colavincenzo:2018cgf}. One could further reduce the computational costs and enhance the precision in this estimation step through various statistical techniques, e.g. the shrinkage method for combining empirical estimates and theoretical models \citep{Pope:2007vz} and the covariance tapering method \citep{Kaufman:2008xx,Paz:2015kwa}.

However, there are multiple caveats to using an estimated covariance matrix. As noted by \cite{Hartlap:2006kj} and known in statistics \citep[see e.g.][]{Anderson:2003xx}, the inverse of an unbiased covariance matrix estimate is biased with respect to the true precision (inverse covariance) matrix that appears in the likelihood function. Therefore, one needs to include a multiplicative correction that is dependent on the data dimension and the number of catalogue samples used for estimation. Further, the error in covariance estimation must be properly propagated to inferred parameter uncertainties, and achieving desired precision requires a large number of catalogue samples \citep{Dodelson:2013uaa,Taylor:2013xx,Percival:2013sga}. Rather than correcting the derived errors, one could adopt the Bayesian principle and treat both the unknown underlying covariance matrix and its estimate as random variables, and marginal{\iz}e over the former given the latter using Bayes' theorem \citep{Sellentin:2015waz}. In addition, as mock catalogues are computationally expensive, they are often produced at fixed fiducial cosmological parameters, whereas in reality the covariance matrix may have cosmological dependence and thus has to vary with the model being tested (see e.g. \citealp{Eifler:2008gx}, in the context of cosmic shear). Failure to account for any of these factors could adversely impact parameter estimation, which should be an important concern to future surveys possessing greater statistical power.

Another fundamental issue to be addressed is the normality assumption itself when the premises of the central limit theorem are not fulfilled. In this scenario, an arbitrarily precise covariance matrix estimate is not sufficient for accurate parameter inference due to significant higher moments in the non-Gaussian likelihood, as demonstrated by \cite{Sellentin:2017fbg} in the context of weak lensing. This also happens to galaxy-clustering measurements on the largest survey scales, where fewer overdensity modes are available due to the finite survey size; if one is to infer parameters sensitive to these large-scale measurements while assuming a Gaussian likelihood, the parameter estimates are likely to be erroneous.

Indeed, in the past there have been efforts to go beyond the Gaussian likelihood approximation in various contexts \citep[e.g. recently][]{Sellentin:2017aii,Seljak:2017rmr}. Motivated by constraints imposed by non-negativity of the power spectrum, \cite{Schneider:2009}, \cite{Keitel:2011} and \cite{Wilking:2013goa} have found a transformation that improves the normality assumption for the bounded configuration-space correlation function which has a non-normal distribution. \cite{Sun:2013nna} have considered the gamma distribution for the power spectrum multipoles and the log-normal plus Gaussian approximation, and assessed their effects on parameter estimation in comparison with the Gaussian approximations. \cite{Kalus:2015lna} have investigated this problem for the three-dimensional power spectrum by deriving the probability distribution of a single-mode estimator for Gaussian random fields, and comparing it with approximations inspired by similar studies for the cosmic microwave background \citep[e.g.][]{Verde:2003ey,Percival:2006ss,Smith:2005ue,Hamimeche:2008ai}. However, these previous works are either limited to the univariate or bivariate distribution, or have neglected window functions due to survey geometry and selection effects, which could correlate independent overdensity modes.

In this work, we focus on the power spectrum monopole, following the Feldman--Kaiser--Peacock approach (\citealp{Feldman:1993ky}, hereafter \citetalias{Feldman:1993ky}). Our work can also be trivially extended to the second and fourth power-law moments measured using the Yamamoto estimator \citep{Yamamoto:2005dz}. The two approaches are equivalent for the monopole moment. Alternatives to using these estimators on large scales would be to either directly fit the observed overdensity modes or to perform an analysis based on the quadratic maximum likelihood (QML) method \citep{Tegmark:1996qt,Tegmark:1997yq}, which would be more optimal given the large-scale window effects in the power spectrum, but computationally more demanding due to evaluation of the large data inverse covariance matrix.

For power-spectrum monopole measurements made using the \citetalias{Feldman:1993ky} procedure, we derive the underlying non-normal probability distribution for the \emph{windowed} galaxy-clustering power spectrum in the linear regime with \emph{random} shot noise. The multivariate normal distribution is reinstated through a Gaussian{\iz}ing transformation that improves data normality, and cosmological dependence of the covariance matrix is fully included by the variance--correlation decomposition. Note that our transformation predicts a new variable whose expectation is not the same transformation of the model. Instead it is simply designed to give a likelihood that is multivariate Gaussian in the data vector. The work is presented as follows.
\begin{enumerate}
    \item We review the \citetalias{Feldman:1993ky} framework for analysing galaxy-clustering measurements in \autoref*{sec:analysis}, and derive the probability distribution of the windowed power spectrum for a Gaussian random field.
    \item A Gaussian{\iz}ation scheme is presented in \autoref*{sec:methodology}, which gives a new power-spectrum likelihood approximation with both cosmological dependence and random scatter in the estimated covariance matrix taken into consideration.
    \item We numerically test our procedure and demonstrate its superiority to the traditional likelihood treatments in \autoref*{sec:tests} by performing inference on the local non-Gaussianity parameter $f_\NL$ and comparing the shape of the new likelihood with that of the true likelihood from simulated data sets.
    \item A simple pipeline is provided in \autoref*{sec:application} for straightforward application of this method. We discuss in \autoref*{sec:conclusion} the applicability of this new approach and motivate future work.
\end{enumerate}
We also provide a summary of notations (Table~\ref{tab:notations}) used in this work, which may be of particular use to the pipeline detailed in \autoref*{sec:application}.
\begin{table*}
    \centering
    \caption{Notations used in the proposed final likelihood analysis pipeline.}
    \setlength{\tabcolsep}{10pt}
    \bgroup
    \def\arraystretch{1.2}
    \begin{tabular}{rll}
        \toprule[0.6pt]
        Symbol & Meaning & Remarks \\
        \midrule[0.4pt]
        $\thetap$ & Cosmological parameter(s) & -- \\
        $\operatorname{\vdot}^{(\fid)}$ & Quantities at the fiducial cosmology & as a superscript in \autoref*{sec:application} \\
        $\operatorname{\vdot}^{(\data)}$ & Measurements/data real{\iz}ations & as a superscript in \autoref*{sec:application} \\
        $N_\simu \equiv m + 1$ & Mock catalogue sample size & -- \\
        $P(\vk)$ & Power spectrum model & -- \\
        $\Pb(k)$ & Band power spectrum model & -- \\
        $a=1,\dots,p$ & Index for $k$-bins & -- \\
        $(R_a,\eta_a)$ & Gamma distribution shape--scale parameters & varying with the cosmological model \\
        $\nu_a$ & Transformation parameter & fixed at $\nu = 1/3$ for simplicity \\
        $\Z$ & Gaussian{\iz}ed data vector & -- \\
        $\mu_a, \sigma^2_a$ & Mean and variance for Gaussian{\iz}ed band power & varying with the cosmological model \\
        $\est{\mSig}$ & Estimated covariance matrix for Gaussian{\iz}ed data & rescaled with varying cosmology \\
        $\Like(\thetap)$ & Likelihood function & -- \\
        \bottomrule[0.6pt]
    \end{tabular}
    \egroup
    \label{tab:notations}
\end{table*}

\section{\texorpdfstring{P\hspace{1.10pt}o\hspace{1.10pt}w\hspace{1.10pt}e\hspace{1.10pt}r\hspace{1.10pt} S\hspace{1.10pt}p\hspace{1.10pt}e\hspace{1.10pt}c\hspace{1.10pt}t\hspace{1.10pt}r\hspace{1.10pt}u\hspace{1.10pt}m\hspace{1.10pt} A\hspace{1.10pt}n\hspace{1.10pt}a\hspace{1.10pt}l\hspace{1.10pt}y\hspace{1.10pt}s\hspace{1.10pt}i\hspace{1.10pt}s}{Power Spectrum Analysis}}
\label{sec:analysis}

\subsection{Galaxy-clustering  measurements in a finite-sized survey}

We assume that the galaxy redshifts measured in a given survey have been converted to comoving distances using a fiducial cosmological model at fixed cosmological parameter(s) $\thetap = \thetap_\fid$. This is required to measure the comoving local galaxy density, and hence the comoving power spectrum. Rather than recalculating the power spectrum for each model to be tested, we include the cosmological model dependence of this translation in the model to be fitted to the data.

Let $n_\gal(\vr)$ be the observed galaxy number density field and $n_\simu(\vr)$ be the number density field in an unclustered random catalogue with expectation $\nbar(\vr) = \biga{n_\gal(\vr)} = \alpha \biga{n_\simu(\vr)}$, where $\alpha$ matches the observed and catalogue mean densities. Following \citetalias{Feldman:1993ky}, one can define the zero-mean field
    \begin{equation}
        \est{F}(\vr) = \frac{w(\vr)}{\sqrt{I}} \qty[n_\gal(\vr) -
        \alpha n_\simu(\vr)]
        \label{eq:weighted field}
    \end{equation}
for a weighted mask $w(\vr)$, where the normal{\iz}ation constant is
    \begin{equation}
        I = \int \dd[3]{\vr} w(\vr)^2 \nbar(\vr)^2 \,.
        \label{eq:field normalisation}
    \end{equation}

The underlying galaxy-clustering power spectrum $P_\true(\vk)$ is equivalent to the Fourier transform of the configuration-space two-point correlation $\xi(\vr)$ by the Wiener--Khinchin theorem \citep[see e.g.][]{Gabrielli:1980xx}. This power spectrum is convolved (denoted with a tilde) with a window due to the mask,
    \begin{equation}
        \widetilde{P}(\vk) = \biga{\abs\big{\est{F}(\vk)}^2} = \int \frac{\dd[3]{\vq}}{(2\uppi)^3} \abs{\winG(\vk - \vq)}^2 P_\true(\vq) + P_{\shot} \,,
        \label{eq:windowed power spectrum}
    \end{equation}
where the function
    \begin{equation}
        \winG(\vk) = \frac{1}{\sqrt{I}} \int \dd[3]{\vr}
        \e^{\im\vk\vdot\vr} w(\vr) \nbar(\vr) \,,
        \label{eq:window function}
    \end{equation}
and there is an additional scale-independent shot noise power (see \appref{app:shot noise})
    \begin{equation}
        P_{\shot} = \frac{1+\alpha}{V} \int \dd[3]{\vr} \nbar(\vr)^{-1} \,.
        \label{eq:shot noise power}
    \end{equation}

The convolved power spectrum may be expanded in the basis of Legendre polynomials $L_\ell(\Delta)$,
    \begin{equation}
        \widetilde{P}(\vk) = \sum_{\ell = 0}^\infty \widetilde{P}_\ell(k,\Delta) L_\ell(\Delta) \,,
        \label{eq:multipole expansion}
    \end{equation}
where $\Delta$ is cosine of the angle between the wavevector $\vk$ and the line of sight. With $L_0 \equiv 1$, the monopole of the convolved power spectrum is a spherical average over the shell $V_k$ at radius $k$,
    \begin{equation}
        \Pb(k) = \int_{V_k} \frac{\dd[3]{\vk}}{V_k} \widetilde{P}(\vk) = \int \frac{\dd[3]{\vq}}{(2\uppi)^3} \winW(k,\vq) P_\true(\vq) + P_\textrm{shot} \,,
        \label{eq:power spectrum monopole}
    \end{equation}
where we have introduced the window function
    \begin{equation}
        \winW(k,\vq) = \int_{V_k} \frac{\dd[3]{\vk}}{V_k}  \abs{\winG(\vk - \vq)}^2 \,.
        \label{eq:window function monopole}
    \end{equation}
Expanding the power spectrum in multipoles allows the computational demand of the convolution to be reduced \citep{Wilson:2015lup}. For a standard linear power spectrum model that is complete with the first three even power-law moments, the convolution could be rewritten requiring three window matrices whose entries are of the form $\winW_\ell(k,\vq)$. However, in order to keep our equations compact, we retain the more general form and focus on the monopole here.

The windowed power spectrum monopole measured in $p$ bins constitutes the band power data vector
    \begin{equation}
        \tY \equiv \bigs{\Pd(k_a)}_{a=1}^p \,,
    \end{equation}
where a hat denotes a real{\iz}ation or an estimator. We can construct another vector
    \begin{equation}
        \Y \equiv \bigs{\abs{\hat{\den}(\vq_i)}^2}_{i=1}^r
        \label{eq:mode power}
    \end{equation}
that estimates the unconvolved power with shot noise at $r$ discrete wavenumbers, where the galaxy overdensity field estimator is
    \begin{equation}
        \hat{\den}(\vr) = \frac{n_\gal(\vr) - \alpha n_\simu(\vr)}{\nbar(\vr)} = \frac{\sqrt{I}}{w(\vr)\nbar(\vr)} \est{F}(\vr) \,.
        \label{eq:realised overdensity field}
    \end{equation}
The discret{\iz}ed analogue to equation~\eqref*{eq:power spectrum monopole} for the measured power is then
    \begin{equation}
        \tY = \mB \Y \,,
        \label{eq:linear compression}
    \end{equation}
where the window function is encoded in the mixing matrix
    \begin{equation}
        B_{ai} = \winW(k_a,\vq_i) \,,
        \label{eq:compression matrix}
    \end{equation}
which may be suitably normal{\iz}ed so that
    \begin{equation}
        \int \frac{\dd[3]{\vq}}{(2\uppi)^3} \winW(k,\vq) = 1 \; \forall k \quad \Leftrightarrow \quad \sum_i B_{ai} = 1 \; \forall a \,.
        \label{eq:window function normalisation}
    \end{equation}

\subsection{Distribution of an individual band power measurement}
\label{ssec:band power distribution}

In this subsection we consider the probability distribution of a band power measurement in a single bin at scale $k_a$. This will be extended to a multivariate distribution including correlations between bins at different scales in \autoref*{sec:methodology}.

\subsubsection{Exact hypo-exponential distribution}
\label{sssec:exact hypo-exponential distribution}

If $\hat{\den}(\vr)$ is directly drawn from a Gaussian random field, then the square amplitude of a single Fourier overdensity mode provides an estimator $\est{P}_i$ for the unconvolved power $P_i \equiv P(\vq_i)$ that is exponentially distributed with the probability density function (PDF)
    \begin{equation}
        \Prob\!\bigs{\est{P}_i} = P_i^{-1} \exp(-\est{P}_i\big/P_i) \,,
        \label{eq:exponential distribution}
    \end{equation}
and has expectation $\Expc\!\bigs{\est{P}_i} = P_i$ and variance $\Var\!\bigs{\est{P}_i} = P_i^2$.

However, galaxy formation is a discrete point process, meaning that the overdensity field real{\iz}ation $\hat{\den}(\vr)$ is a \emph{Poisson sample} of the underlying Gaussian overdensity field \citep{Peebles:1980xx,Feldman:1993ky}. Consequently, each mode-power estimator $Y_i$ has an additional independent shot noise component $\epsilon_i$,
    \begin{equation}
        Y_i = \est{P}_i + \epsilon_i \,.
    \end{equation}
In the large galaxy number limit, the shot noise $\epsilon_i$ is also exponentially distributed (see \appref{app:shot noise}).

For the bin centred at $k_a$, the window function mixes mutually independent exponential variables $\bigb{\est{P}_i}$ and $\qty{\epsilon_i}$ into the band power
    \begin{equation}
        \wY_a = \sum_{i=1}^{r} B_{ai} Y_i = \sum_{i=1}^{r} B_{ai} \bigp{\est{P}_i + \epsilon_i} \,.
        \label{eq:exponential mixture}
    \end{equation}
This is an exponential mixture that follows the hypo-exponential distribution (see \appref{app:hypo-exponential distribution})
    \begin{equation}
        \Prob\!\bigs{\wY_a;\{\lambda_{i'}\}} = \sum_{i'} \biggp{\prod_{j' \neq i'} \frac{1}{1 - \lambda_{j'}/\lambda_{i'}}} \lambda_{i'}^{-1} \exp(- \lambda_{i'}^{-1} \wY_a) \,,
        \label{eq:hypo-exponential distribution}
    \end{equation}
with positive scale parameters
    \begin{equation}
        \lambda_{i'} \in \{ B_{ai} P_{i}, B_{ai} P_\shot: i = 1,\dots,r \}
        \label{eq:hypo-exponential scale parameters}
    \end{equation}
being the individual contributions of the unconvolved power and shot noise power to the $a$-th bin. It is understood that a well-defined limit is taken in equation~\eqref*{eq:hypo-exponential distribution} in the case $\lambda_{i'} = \lambda_{j'}$.

We now see that the band power measurement $\wY_a = \Pd(k_a)$ is hypo-exponentially distributed for a Poisson-sampled Gaussian overdensity field. By the central limit theorem, as the number of contributing modes in equation~\eqref*{eq:exponential mixture} increases, the hypo-exponential variable $\wY_a$ converges in distribution to a normal variable. This is the basis for the normality assumption often used in power spectrum analyses: the underlying power as well as shot noise becomes normally distributed, with the latter subtracted as a deterministic quantity to recover the former. However, on the largest scales in a survey where the number of overdensity modes is the fewest, there is clear deviation between the normal distribution and the hypo-exponential distribution.

\subsubsection{Gamma distribution approximation}
\label{sssec:gamma distribution approximation}

The univariate PDF given by equation~\eqref*{eq:hypo-exponential distribution} is in a difficult form to manipulate as it involves many uncompressed scale parameters $\{\lambda_{i'}\}$. A robust approximation is the exponentially modified gamma distribution with one shape parameter $R$ and two scale parameters $(\tau,\eta)$ \citep[see][]{Golubev:2016203}, but determining these parameters involves solving a cubic algebraic equation which is cumbersome, making the procedure followed in this paper computationally demanding or even unfeasible. We can adopt a simpler approximation with the gamma distribution
    \begin{equation}
        \Prob_\Gamma\!\bigs{\wY_a;R,\eta} = \frac{\eta^{-R}}{\varGamma(R)} \wY_a^{R-1} \e^{-\wY_a/\eta} \,,
        \label{eq:gamma distribution}
    \end{equation}
where by matching the mean and variance of the hypo-exponential distribution we can easily write down the shape and scale parameters
    \begin{equation}
        \begin{split}
            & R = \Expc\!\bigs{\wY_a}^2 \Big/ \Var\!\bigs{\wY_a} = \biggp{\sum_{i'} \lambda_{i'}}^2\bigg/{\sum_{i'} \lambda_{i'}^2} \,, \\
            & \eta = \Var\!\bigs{\wY_a} \Big/ \Expc\!\bigs{\wY_a} = \biggp{\sum_{i'} \lambda_{i'}^2}\bigg/{\sum_{i'} \lambda_{i'}} \,.
        \end{split}
        \label{eq:gamma distribution parameters}
    \end{equation}
In place of the mean and variance parameters of a normal distribution, these two gamma distribution parameters determine the \emph{non-normal} distribution of the band power measurement in each bin, and have a natural interpretation in our context: the shape $R$ is the effective number of independent overdensity modes contributing to the bin under window function mixing, and the scale $\eta$ is the effective convolved power in that bin. In the limiting case that there is only a single non-vanishing mode (e.g. pure shot noise), the gamma distribution coincides exactly with the hypo-exponential distribution, both of which reduce to the exponential distribution.

In principle, the shape and scale parameters $(R,\eta)$ could be calculated for each bin using analytical expressions of the band power expectation $\Expc\!\bigs{\wY_a}$ and variance $\Var\!\bigs{\wY_a}$, with knowledge of the full window mixing matrix $\mB$ and all observed overdensity modes $\bigb{\hat{\den}(\vq_i)}$. Practically, these quantities are not always available, as fast window convolution is now often performed in configuration space with ever larger surveys and higher resolution requirements, and the windowed power spectrum is computed via a Hankel transform of the convolved two-point correlation multipoles $\widetilde{\xi}_\ell(\vr)$ \citep{Beutler:2016arn,Wilson:2015lup}. Under these circumstances, the variance of the measured band power $\Var\!\bigs{\wY_a}$ may be estimated from mock catalogues, provided the error in this estimation is subdominant compared to other sources of uncertainty. However, it is worth pointing out ongoing efforts in developing accurate analytic covariance matrices that could possibly evade the problem of covariance estimation altogether, e.g. \cite{Li:2018scc}.

\subsubsection{Normal distribution assumption}

For the sake of completeness and for reference, we write down the univariate normal distribution for the band power $\wY_a$ in terms of the shape--scale parameters $(R,\eta)$ for the $a$-th bin,
    \begin{equation}
        \Prob_{\Norm}\!\bigs{\wY_a;R\eta,R\eta^2} = \frac{1}{\sqrt{2\uppi R\eta^2}} \exp[-\frac{\bigp{\wY_a - R\eta}^2}{2 R\eta^2}] \,,
        \label{eq:normal distribution assumption}
    \end{equation}
so that its expectation and variance match those of the exact hypo-exponential and approximate gamma distributions.

\section{\texorpdfstring{G\hspace{1.10pt}a\hspace{1.10pt}u\hspace{1.10pt}s\hspace{1.10pt}s\hspace{1.10pt}i\hspace{1.10pt}a\hspace{1.10pt}n\hspace{1.10pt}i\hspace{1.10pt}z\hspace{1.10pt}a\hspace{1.10pt}t\hspace{1.10pt}i\hspace{1.10pt}o\hspace{1.10pt}n\hspace{1.10pt} a\hspace{1.10pt}n\hspace{1.10pt}d\hspace{1.10pt}\\ V\hspace{1.10pt}a\hspace{1.10pt}r\hspace{1.10pt}i\hspace{1.10pt}a\hspace{1.10pt}n\hspace{1.10pt}c\hspace{1.10pt}e\hspace{1.10pt}--\hspace{1.10pt}C\hspace{1.10pt}o\hspace{1.10pt}r\hspace{1.10pt}r\hspace{1.10pt}e\hspace{1.10pt}l\hspace{1.10pt}a\hspace{1.10pt}t\hspace{1.10pt}i\hspace{1.10pt}o\hspace{1.10pt}n\hspace{1.10pt} D\hspace{1.10pt}e\hspace{1.10pt}c\hspace{1.10pt}o\hspace{1.10pt}m\hspace{1.10pt}p\hspace{1.10pt}o\hspace{1.10pt}s\hspace{1.10pt}i\hspace{1.10pt}t\hspace{1.10pt}i\hspace{1.10pt}o\hspace{1.10pt}n}{Gaussianization and Variance--Correlation Decomposition}}
\label{sec:methodology}

A multivariate PDF transformation for $\Y \xmapsto{\mB} \tY$ is captured by the Jacobian factor $\mathcal{J} = \det(\mB \trans{\mB})$, where the distribution of $\Y$ is the product of independent exponential components. The window mixing matrix $\mB \in \R_+^{p \times r}$ is non-square ($p < r$), and expressing the transformed PDF explicitly in terms of the band power vector $\tY$, whose components are correlated, requires the inversion of $\mB$. One could introduce $(r-p)$ `helper components' in $\tY$ to pad $\mB$ into a square matrix, and eventually marginal{\iz}e out these additional random variables. However, the linear transformation induced by the padded square matrix will generally map the domain of the random vector $\Y$ from $\R_+^r$ to a different domain in $\R^r$, making marginal{\iz}ation difficult and susceptible to the `curse' of dimensionality, which is likely as $r \gg p$ for massive data compression.

Instead of trying to determine a full multivariate transformation for the windowed band power, we subscribe to a \emph{component-wise} Gaussian{\iz}ation strategy, and reinstate the multivariate normality assumption in the \emph{Gaussian{\iz}ed} data vector $\Z \mapsfrom \tY$. The reasoning behind this is two fold. First, each component of the random vector is now certainly univariate normal, as should be the case for a bona fide multivariate normal distribution. Secondly, if the cross-bin correlation is weak and the covariance matrix has a narrow-band structure, the components become approximately independent and univariate Gaussian{\iz}ation is equivalent to multivariate Gaussian{\iz}ation; this can be achieved with a suitable binning choice where the bin width is chosen to be greater than the known window correlation length. Although sophisticated full Gaussian{\iz}ation schemes exist \citep[e.g.][]{Laparra:2011xxx}, they are often iteratively applied based on empirical data and may not be suitable for forward modelling of theoretical models. We thus leave more advanced multivariate Gaussian{\iz}ation methods for future investigations.

We propose a simple Gaussian{\iz}ation scheme using the Box--Cox transformation. This scheme has already been applied in cosmology to deal with non-Gaussian parameter spaces \citep{Joachimi:2011xxx,Schuhmann:2015dma}, but our context and implementation differ from these studies. An alternative scheme directly transforming a non-normal distribution into the standard normal distribution exists by matching the cumulative distribution function; however, computationally costly numerical integration is necessary for calculating transformed moments, so we relegate this scheme to \appref{app:error-function transformation} for reference. As we shall see later, the Box--Cox transformation suppresses higher order moments to achieve \emph{approximate} Gaussian{\iz}ation and is sufficiently accurate for our purposes. To perform the transformation, we use the fiducial shape--scale parameters $(R_\fid,\eta_\fid)$ which are determined by the power spectrum model $P_\fid(k)$ at fixed cosmological parameter(s) $\thetap = \thetap_\fid$.

\subsection{Box--Cox transformation}
\label{ssec:Box--Cox transformation}

We define the Box--Cox transformation \citep{Box:1964xxx} for each component of $\Y$ (index suppressed for brevity) by
    \begin{equation}
        Z = \wY^{\nu} \,, \quad \nu > 0 \,,
        \label{eq:Box--Cox transformation}
    \end{equation}
where positive $\nu$ is chosen to ensure regularity. The transformed PDF in $Z$ is now
    \begin{equation}
        \Prob\!\bigs{Z\bigp{\wY};R,\eta} = \big\vert{\mathcal{J}_Z\bigp{\wY}}\big\vert^{-1} \frac{\eta^{-R}}{\varGamma(R)} \wY^{R-1} \e^{-\wY/\eta} \,,
    \end{equation}
where $\mathcal{J}_Z\bigp{\wY} = \nu \wY^{\nu-1}$ is the Jacobian for the transformation $\wY \mapsto Z$. The transformed $K$\textsuperscript{th} moment is given by
    \begin{equation}
        \Expc\!\bigs{Z^K} = \frac{\varGamma(R+K\nu)}{\varGamma(R)} \eta^{K\nu} \,,
    \end{equation}
and we can write down the Gaussian{\iz}ed mean and variance
    \begin{equation}
        \begin{split}
            & \mu(R,\eta) \equiv \Expc[Z] = \frac{\varGamma(R + \nu)}{\varGamma(R)} \eta^\nu \,, \\
            & \sigma^2(R,\eta) \equiv \Var[Z] = \frac{\varGamma(R + 2\nu) \varGamma(R) - \varGamma(R + \nu)^2}{\varGamma(R)^2} \eta^{2\nu} \,.
        \end{split}
        \label{eq:transformed mean and variance}
    \end{equation}

To determine the transformation parameter $\nu$, we demand the third central moment vanish for the \emph{fiducial} model parameters $\thetap_\fid$,
    \begin{align}
        0 &= \Expc\!\bigs{Z^3} - 3 \Expc\!\bigs{Z^2} \Expc\!\bigs{Z} + 2 \Expc\!\bigs{Z}^3 \nonumber \\
        &%
        \begin{multlined}[b][0.85\linewidth]
            = \bigg\lbrace \frac{\varGamma(R_\fid + 3\nu)}{\varGamma(R_\fid)} - 3 \frac{\varGamma(R_\fid + 2\nu)}{\varGamma(R_\fid)} \frac{\varGamma(R_\fid + \nu)}{\varGamma(R_\fid)} \\ + 2 \qty[\frac{\varGamma(R_\fid + \nu)}{\varGamma(R_\fid)}]^3 \bigg\rbrace \eta^{3\nu} \,.
        \end{multlined}
        \label{eq:optimal transformation parameter}
    \end{align}

\begin{figure}
    \centering
    \includegraphics[width=0.975\linewidth]{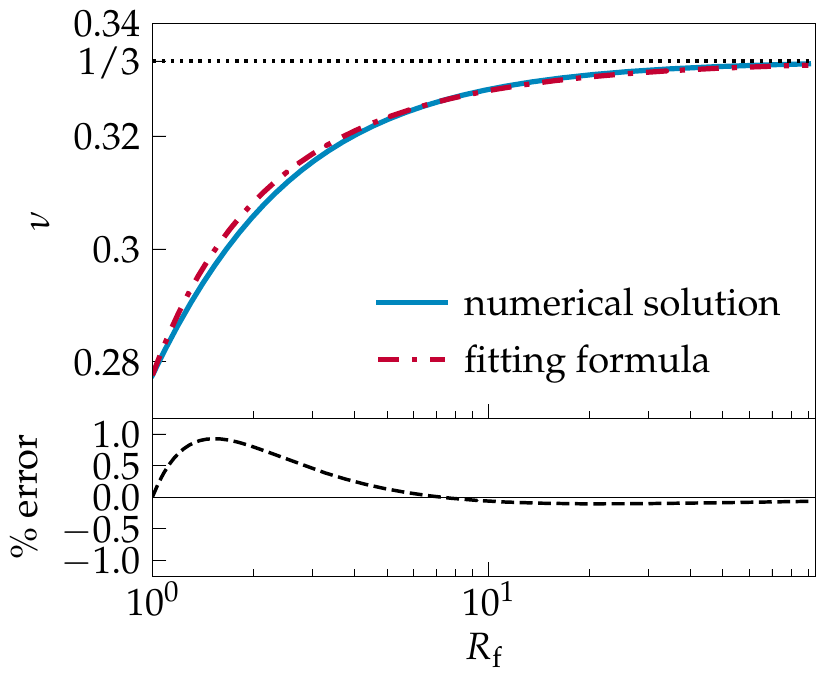}
    \caption{The dependence of the optimal Box--Cox transformation parameter, $\nu$, on the fiducial gamma-distribution shape parameter $R_\fid$. \textit{Top panel}: comparison of the numerical solution to equation~\eqref*{eq:optimal transformation parameter} (\textit{solid line}) with the empirical fitting formula given by equation~\eqref*{eq:optimal transformation-parameter fitting formula} (\textit{dash--dotted line}) for determining the optimal transformation parameter $\nu$ as a function of the fiducial shape parameter $R_\fid$. \textit{Bottom panel}: the percentage error of the fitting formula from the numerical solution (\textit{dashed line}) is at the sub per cent level.}
    \label{fig:fitting formula}
\end{figure}
The dependence of $\nu$ on $R_\fid$ required to satisfy this constraint is shown in Fig.~\ref{fig:fitting formula} as a numerical solution (dashed blue line). The observed asymptotic behaviour can be understood by considering the expansion of gamma function ratios \citep{Burić:2011}
    \begin{multline}
        \frac{\varGamma(A+b)}{\varGamma(A)} \sim A^b \Big[1 + \frac{(b-1)b}{2} A^{-1} \\ + \frac{(3b-1)(b-2)(b-1)b}{24} A^{-2} + \cdots\Big] \,,
    \end{multline}
so as $R_\fid \to \infty$ with $\eta R_\fid < \infty$ (finite mean), equation~\eqref*{eq:optimal transformation parameter} becomes
    \begin{equation}
        0 \simeq R_\fid^{-2} \nu^3 (3\nu-1) \,.
    \end{equation}
The non-trivial solution is $\nu = 1/3$. The precise value of $\nu$ matters less for increasing $R_\fid$ owing to the suppression factor $R_\fid^{-2}$, a manifestation of asymptotic normality in the limit $R_\fid \gg 1$.

An empirical fitting formula for the solution to equation~\eqref*{eq:optimal transformation parameter} is given by
    \begin{equation}
        \nu \approx \frac{1}{3} + \num{0.042} \qty[1 - \exp(\num{0.85}/R_{\fid}^{\num{0.85}})] \,.
        \label{eq:optimal transformation-parameter fitting formula}
    \end{equation}
Comparing this fit to the true solution in Fig.~\ref{fig:fitting formula} shows that this fitting formula performs well, being accurate to sub per cent levels. As we shall see in \autoref*{sec:tests}, in fact simply assuming the fixed value $\nu = 1/3$ works very well in realistic situations; our Gaussian{\iz}ation scheme is robust to variation in $\nu$ with the fiducial parameter $R_\fid$, i.e. the choice of fiducial cosmology.

\subsection{Covariance treatment}
\label{ssec:covariance treatment}

In this subsection we discuss two general treatments of estimated covariance matrices which could be in the band power data $\tY$ or in the Gaussian{\iz}ed data $\Z$, although our treatments are applied to the Gaussian{\iz}ed data in \autoref*{sec:application}.

\subsubsection{Cosmological parameter dependence}

Now that we have a Gaussian{\iz}ing transformation, the model dependence of the covariance matrix still needs to be considered. This sub-subsection shows how the parameter dependence can be included in the covariance matrix estimate analytically. In \autoref*{sec:analysis}, we have ignored the integral constraint in the modelling of the power spectrum, which biases the power spectrum measured due to estimation of the mean galaxy number density from the galaxy catalogue itself; since this offset in the measured power can be subtracted \citep[see e.g.][]{Peacock:1991xx,Beutler:2013yhm}, we do not consider its contribution here in this work.

A generic covariance matrix $\mSig$ may be decomposed into the diagonal matrix $\mLamb = (\Diag\mSig)^{1/2}$ and the correlation matrix $\mC$,
    \begin{equation}
        \mSig = \mLamb \mC \mLamb \,,
        \label{eq:covariance decomposition}
    \end{equation}
where $\mLamb^2$ is the diagonal matrix of the variances. For the band power spectrum on large scales, \emph{whether Gaussian{\iz}ed or not}, the off-diagonal correlation in $\mC$ is solely induced by the window function as encoded in the mixing matrix $\mB$. Whilst this mixing matrix does depend on the fiducial cosmological model through the distance--redshift relation, crucially it does not depend on the model being tested through the power spectrum.\footnote{\!\!\!A caveat is that, in principle, there could be redshift-space effects that render the overall window function dependent on the underlying cosmology, e.g. a survey boundary defined by a maximum redshift.} This insight makes the variance--correlation decomposition particularly useful, for the decomposition into $\mLamb$ and $\mC$ is precisely the separation of any cosmology dependence from cosmology independence in the covariance matrix $\mSig$. Therefore one may obtain a covariance matrix estimate $\est{\mSig}_\fid = \mLamb_\fid \mC \mLamb_\fid$ from mock catalogues produced at a fixed cosmology $\thetap = \thetap_\fid$ with the fiducial power spectrum model $P_\fid(k)$, and calibrate this estimate by rescaling with the diagonal variances to allow for varying cosmology,
    \begin{equation}
        \est{\mSig}(\thetap) = \mLamb(\thetap) \mLamb_\fid^{-1} \est{\mSig}_\fid \mLamb_\fid^{-1} \mLamb(\thetap) \,.
        \label{eq:covariance calibration}
    \end{equation}
For instance, for the Gaussian{\iz}ed band power $\Z$  at cosmological parameter(s) $\thetap$, the entries in the diagonal variance matrix are
    \begin{equation}
        \Diag\mLamb^2(\thetap) = \Var[\Z] = \bigs{\sigma_a^2(R_a,\eta_a)}_{a=1}^p
        \label{eq:rescaling matrix}
    \end{equation}
as given by equation~\eqref*{eq:transformed mean and variance}, where the gamma shape--scale parameters $(R_a,\eta_a)$ for each bin depend on cosmological parameter(s) $\thetap$ through the power (see equations~\ref*{eq:hypo-exponential scale parameters} and \ref*{eq:gamma distribution parameters}).

\subsubsection{Covariance matrix estimates as random variables}

Covariance estimation from mock catalogues gives an inherently random quantity. Instead of directly substituting the estimated covariance in a probability distribution, a more principled approach is to marginal{\iz}e out the unknown underlying covariance matrix using Bayes' theorem (\citealp{Sellentin:2015waz}, hereafter \citetalias{Sellentin:2015waz}). Let $\est{\mSig}_\fid$ be an unbiased covariance matrix estimate calculated from $N_\simu \equiv m + 1$ samples of the data vector from mock catalogues at the fiducial cosmology. Using the uninformative Jeffreys prior on the unknown true covariance $\mSig_\fid$,
    \begin{equation}
        \Prior[\mSig_\fid] \propto \abs{\mSig_\fid}^{-(p+1)/2} \,,
    \end{equation}
and the fact that an empirical covariance matrix estimate $\est{\mSig}_\fid$ has the Wishart distribution conditional on $\mSig_\fid$,
    \begin{equation}
        \Prob\!\bigs{\bgiven{\est{\mSig}_\fid}{\mSig_\fid}} \propto \frac{\big\vert{\est{\mSig}_\fid}\big\vert^{(m-p-1)/2}}{\abs{\mSig_\fid}^{m/2}} \exp[-\frac{m}{2} \tr(\mSig_\fid^{-1} \est{\mSig}_\fid)] \,,
    \end{equation}
one could show that the posterior distribution of $\bgiven{\mSig_\fid}{\est{\mSig}_\fid}$ is inverse Wishart with the PDF
    \begin{equation}
        \Prob\!\bigs{\bgiven{\mSig_\fid}{\est{\mSig}_\fid}} \propto \frac{\big\vert{\est{\mSig}_\fid}\big\vert^{m/2}}{\abs{\mSig_\fid}^{(m+p+1)/2}} \exp[-\frac{m}{2} \tr(\est{\mSig}_\fid \mSig_\fid^{-1})] \,.
    \end{equation}
It is this distribution that one needs to marginal{\iz}e over to replace the unknown true covariance $\mSig_\fid$ with its estimate $\est{\mSig}_\fid$. The same derivation follows exactly for the rescaled covariance estimate $\est{\mSig}(\thetap)$, since our decomposition does not affect the \citetalias{Sellentin:2015waz} marginal{\iz}ation procedure (see \appref{app:covariance marginalisation}).

In addition to the covariance matrix estimate for the Gaussian{\iz}ed data $\Z$, if one estimates the gamma distribution parameters $\{R_a\}$ and $\{\eta_a\}$ (see equation~\ref*{eq:gamma distribution parameters}) from the empirical covariance matrix estimate of the band power data $\tY$, then these parameters should also be considered as random variables; in particular, the shape adopted for the likelihood is itself an estimated quantity, and this could in principle also bias the recovered likelihood. In \autoref*{sec:tests}, we will see that in practice this is not an issue: for estimated covariance matrices with significant uncertainties, the effect of the covariance estimation (allowed for by the \citetalias{Sellentin:2015waz} marginal{\iz}ation procedure) itself dominates. For estimated covariance matrices with low noise, the two effects can become comparable, but in this situation the size of both effects is small.

\subsection{Likelihood form}

When the mean vector $\vmu$ for the Gaussian data vector $\Z$ and its covariance matrix $\mSig$ are known exactly, the multivariate normal PDF simply reads
    \begin{equation}
        \Prob_{\Norm}[\Z;\vmu,\mSig] = \big\vert{2\uppi\mSig}\big\vert^{-1/2} \exp[-\frac{1}{2} \chi^2(\Z;\vmu,\mSig)] \,,
        \label{eq:multivariate normal distribution}
    \end{equation}
where the quantity
    \begin{equation}
        \chi^2(\Z;\vmu,\mSig) \equiv \trans{(\Z - \vmu)} {\mSig}^{-1} (\Z - \vmu) \,.
      \label{eq:chi-square}
    \end{equation}
This is the multivariate version of equation~\eqref*{eq:normal distribution assumption} but now for the \emph{Gaussian{\iz}ed} band power $\Z$. The \citetalias{Sellentin:2015waz} procedure replaces the underlying covariance $\mSig$ with an estimate $\est{\mSig}$, and changes this to a modified $t$-distribution (see \appref{app:covariance marginalisation})
    \begin{equation}
        \Prob_t\!\bigs{\Z;\vmu,\est{\mSig},m} = c_p \big\vert{\est{\mSig}}\big\vert^{-1/2} \qty[1 + \frac{\chi^2\bigp{\Z;\vmu,\est{\mSig}}}{m}]^{-(m+1)/2} \,,
        \label{eq:modified t-distribution}
    \end{equation}
where the normal{\iz}ation constant is
    \begin{equation}
        c_p = (m\uppi)^{-p/2} \frac{\varGamma[(m+1)/2]}{\varGamma[(m-p+1)/2]} \,.
    \end{equation}
The PDF given in equation~\eqref*{eq:modified t-distribution}, when regarded as a function of cosmological parameter(s) $\thetap$ through the dependence of $\vmu$ and $\est{\mSig}$ on the power spectrum model as functions of $\{(R_a,\eta_a)\}$,
    \begin{equation}
        \Like\!\bigp{\thetap;\Z,\est{\mSig}_\fid,m} = \Prob_t\!\bigs{\Z;\vmu(\thetap),\est{\mSig}(\thetap),m} \,,
        \label{eq:full likelihood}
    \end{equation}
is a key result of this work.

In the next section, we shall test different aspects of our procedure used to derive the new likelihood, using Monte Carlo simulations matched to the specifications of future survey data. This leads us to formulating a simple pipeline for likelihood analysis of galaxy surveys, which we present in \autoref*{sec:application}.

\section{\texorpdfstring{T\hspace{1.10pt}e\hspace{1.10pt}s\hspace{1.10pt}t\hspace{1.10pt}i\hspace{1.10pt}n\hspace{1.10pt}g\hspace{1.10pt} w\hspace{1.10pt}i\hspace{1.10pt}t\hspace{1.10pt}h\hspace{1.10pt} N\hspace{1.10pt}u\hspace{1.10pt}m\hspace{1.10pt}e\hspace{1.10pt}r\hspace{1.10pt}i\hspace{1.10pt}c\hspace{1.10pt}a\hspace{1.10pt}l\hspace{1.10pt} S\hspace{1.10pt}i\hspace{1.10pt}m\hspace{1.10pt}u\hspace{1.10pt}l\hspace{1.10pt}a\hspace{1.10pt}t\hspace{1.10pt}i\hspace{1.10pt}o\hspace{1.10pt}n\hspace{1.10pt}s}{Testing with Numerical Simulations}}
\label{sec:tests}

In order to test our proposed likelihood form, we create Monte Carlo simulations of data vectors by generating exponentially distributed overdensity mode power and shot noise, and then convolve with a chosen survey window before extracting the measured power spectra. Given the survey volume of DESI and \textit{Euclid}, we consider scales $k = \numrange[range-phrase=\text{--}]{1.58e-3}{1.58e-2}\si{\h\per\mega\parsec}$ covering an order of magnitude, and select overdensity wavenumbers $\vq$ using an inverse-volume distribution $\Prob[\vq] \propto \abs{\vq}^3$. The chosen window function has a Gaussian shape with full width at half-maximum (FWHM) \SI{1.88e-3}{\h\per\mega\parsec}. We divide the scales into $p = 9$ bins so that the cross-bin correlation is weak under this window function. Given an input cosmology, the power spectrum model is specified as follows:
\begin{enumerate}
    \item The underlying galaxy power spectrum is calculated using the fitting formula for the matter transfer function $T(k)$ by \cite{Eisenstein:1997ik}, with the large-scale galaxy linear bias fixed at $b_0 = 1.87$ and other cosmological parameters set to \textit{Planck} 2018 values \citep{Aghanim:2018eyx};
    \item The shot noise power is calculated using equation~\eqref*{eq:shot noise power} with a pessimistic number density $\nbar = \SI{5e-4}{\cubic\h\per\cubic\mega\parsec}$. The number densities predicted for DESI and \textit{Euclid} are higher \citep[e.g.][]{Duffy:2014lva}, so our analysis is conservative with respect to the effect of shot noise on large scales.
\end{enumerate}

\subsection{Testing distribution normality}

\begin{figure}
    \centering
    \includegraphics[width=0.975\linewidth]{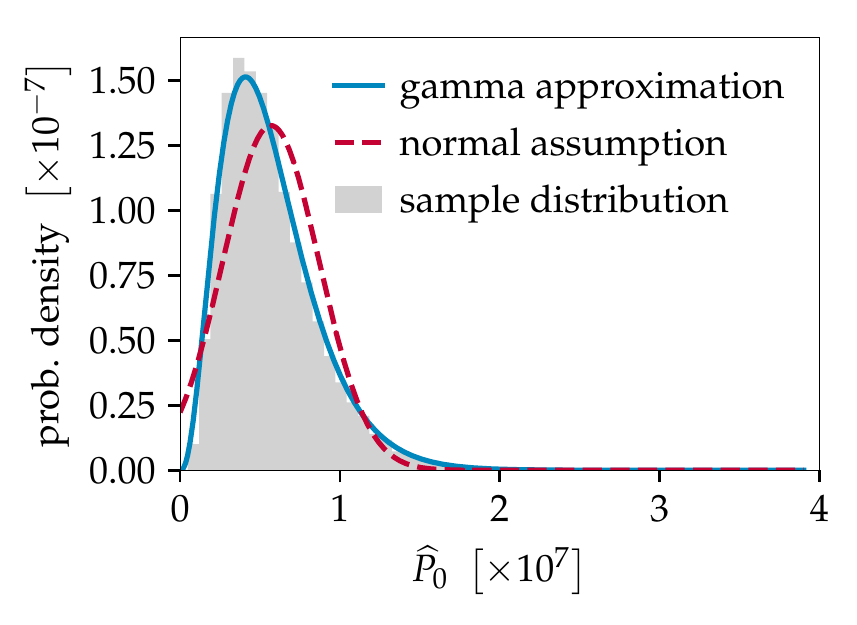}
    \caption{Distribution of the band power spectrum in the lowest-$k$ bin, centred at $k\approx \SI{0.0024}{\h\per\mega\parsec}$. The effective number of independent modes in this bin is $R \approx 4$. We compare the exact hypo-exponential distribution (\textit{filled region}) sampled from \num{40000} real{\iz}ations, the gamma distribution approximation (\textit{solid line}) and the normal distribution assumption with the same mean and variance (\textit{dashed line}) for the band power measurement.}
    \label{fig:band power distribution}
\end{figure}

In Fig.~\ref{fig:band power distribution}, we compare the sampled hypo-exponential distribution (equation~\ref*{eq:hypo-exponential distribution}), the gamma distribution approximation (equation~\ref*{eq:gamma distribution}) and the normal distribution assumption (equation~\ref*{eq:normal distribution assumption}) with the same mean and variance for the band power in the bin centred at $k \approx \SI{0.0024}{\h\per\mega\parsec}$, which has an effective number of independent modes $R \approx 4$.
The assumed normal distribution has a peak shifted from the underlying hypo-exponential distribution; on the other hand, the gamma distribution is a good approximation that matches both the peak and the tails well.

One may wish to quantify the improvement in \emph{multivariate} normality our component-wise Gaussian{\iz}ation could bring to the band power data vector. A key defining property of a multivariate normal variable $\X$ is that any projection $\X \mapsto \trans{\vb*{t}} \X \in \R$, for some vector $\vb*{t}$, gives a univariate normal variable. Hence as a simple multivariate normality test, given a set of samples $\{ \vx_i \}$ for $\X$, one could randomly choose some directions $\vb*{t}$ and perform univariate normality tests on the projected samples $\{ \trans{\vb*{t}} \vx_i \}$ \citep[see e.g.][]{Shao:2010}.

We perform the D'Agostino--Pearson normality test \citep{D'Agostino:1973} on \num{10000} random projections of \num{40000} samples of the band power vector $\tY$, which returns the $p$-value that character{\iz}es how extreme the sample real{\iz}ations are under the null hypothesis that the underlying distribution were indeed normal. It must be emphas{\iz}ed here that the $p$-value itself is not a meaningful indicator of normality, as it varies depending on the sample size; rather it is the comparison of the $p$-values with the same sample that signifies relative departure from normality. We find $p = \num{0.01}$ for $\tY$ without Gaussian{\iz}ation; with Gaussian{\iz}ation $\tY \mapsto \Z$, however, we find improved $p = \num{0.08}$ given these samples.

\subsection{Testing covariance treatment}

\begin{figure*}
    \centering
    \includegraphics[width=0.975\linewidth]{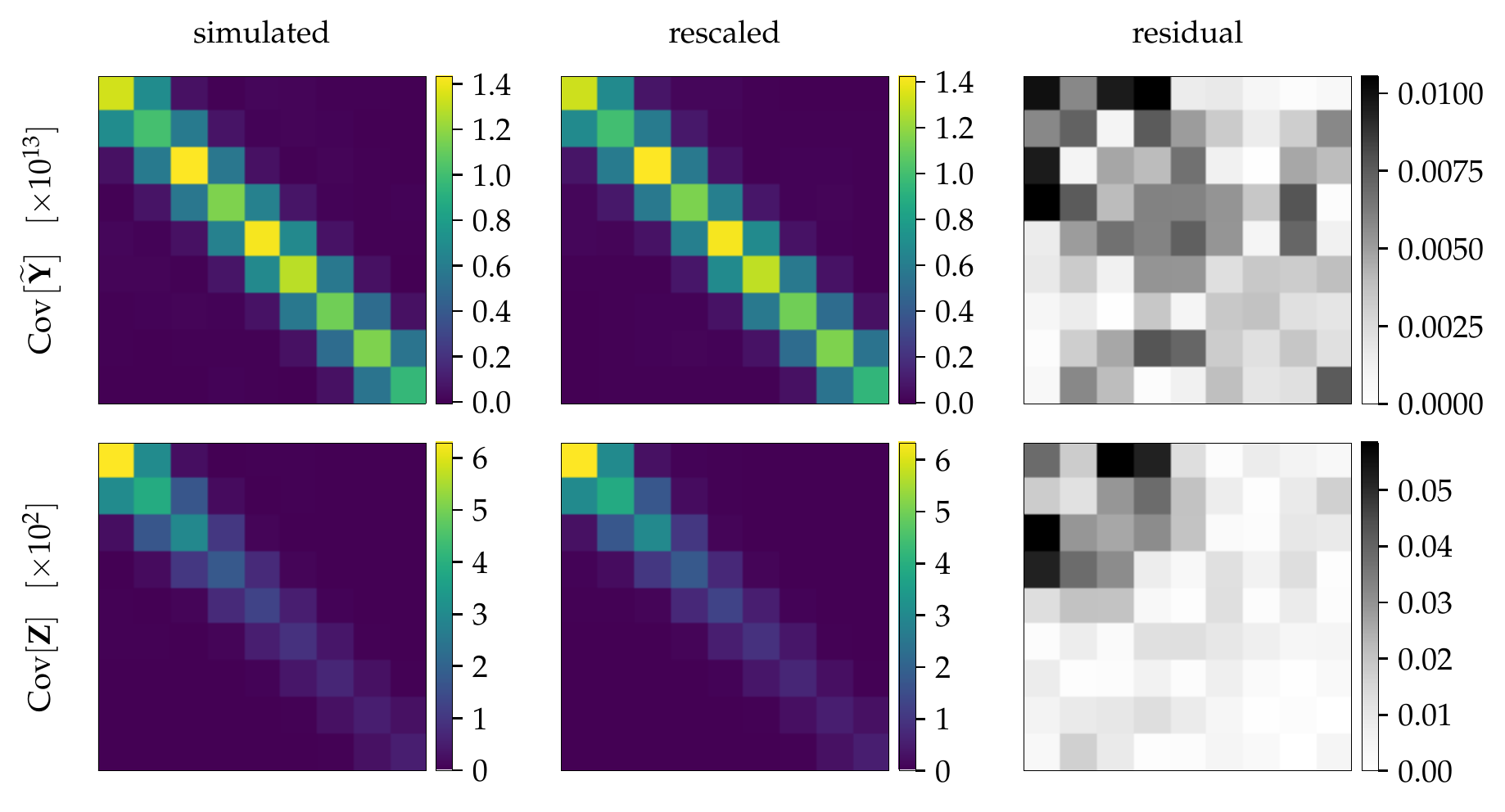}
    \caption{Validation of the variance--correlation decomposition to account for the parameter dependence of the covariance matrix by comparing directly sampled covariance matrices with rescaled covariance matrix estimates. \textit{Left-hand panel}: covariance matrices sampled at a `true' cosmology input ($H_0 = \num{67.4}$); \textit{middle panel}: covariance estimates at the `fiducial' cosmology ($H_0 = \num{73.2}$) rescaled using equation~\eqref*{eq:covariance calibration} to match the `true' cosmology; \textit{right-hand panel}: residuals (element-wise absolute differences) between the covariance matrices in the left-hand and middle panels. The \textit{top row} and the \textit{bottom row} are produced from \num{40000} real{\iz}ations of the band power $\tY$ and the Gaussian{\iz}ed band power $\Z$, respectively.}
    \label{fig:covariance calibration}
\end{figure*}

To test the variance--correlation decomposition proposed in Section~\ref*{ssec:covariance treatment}, we generate one set of \num{40000} band power data real{\iz}ations with the Hubble parameter set to $H_0 = \num{67.4}$, and an additional set of \num{40000} real{\iz}ations generated with $H_0 = \num{73.2}$. The former set gives a sampled `true' covariance matrix $\est{\mSig}$, and the latter gives a `fiducial' covariance estimate $\est{\mSig}_\fid$ which is then rescaled using equation~\eqref*{eq:covariance calibration} to match the `true' cosmology. For both band power $\tY$ without Gaussian{\iz}ation and Gaussian{\iz}ed band power $\Z$, Fig.~\ref{fig:covariance calibration} shows that the differences between the directly sampled `true' covariance matrices and the rescaled covariance estimates are small; this validates the decomposition as a means to include cosmological dependence of the covariance matrix.

Since covariance marginal{\iz}ation and decomposition do not mutually affect each other, we do not test the effect of the \citetalias{Sellentin:2015waz} procedure (the modified-$t$ likelihood) separately here but point the reader to \cite{Sellentin:2015waz} and \cite{Sellentin:2017fbg} for reference.

\subsection{Testing likelihoods for parameter inference}

The ultimate goal of power-spectrum likelihood analysis is to constrain cosmological parameters, so the primary aim of our numerical simulations is to test our new likelihood function after Gaussian{\iz}ation and covariance rescaling.

To summar{\iz}e, we have proposed the following steps for deriving the new likelihood function~\eqref*{eq:full likelihood}, which we now tweak to isolate their effects:
\begin{enumerate}
    \item \textit{Gaussian{\iz}ation} -- the data vector can remain un-Gaussian{\iz}ed $\tY$, Gaussian{\iz}ed $\Z_{1/3}$ with a fixed parameter $\nu = 1/3$, or Gaussian{\iz}ed $\Z_\nu$ with a fitted transformation parameter $\nu$ given by the formula in equation~\eqref*{eq:optimal transformation-parameter fitting formula};
    \item \textit{Covariance rescaling} -- the fiducial covariance matrix estimate is either fixed at $\est{\mSig}_\fid$ when calculating the likelihood, or rescaled to $\est{\mSig}(\thetap)$ using equation~\eqref*{eq:covariance calibration} to account for parameter dependence;
    \item \textit{Covariance marginal{\iz}ation} -- the Hartlap-debiased precision matrix estimate \citep{Hartlap:2006kj} can either be substituted directly into the Gaussian likelihood $\Like_\textrm{G}(\thetap)$ (see equation~\ref*{eq:multivariate normal distribution}), or the \citetalias{Sellentin:2015waz} marginal{\iz}ation procedure can be used to give the modified-$t$ likelihood $\Like_t(\thetap)$ (see equation~\ref*{eq:modified t-distribution}).
\end{enumerate}
Different combinations of these choices give the likelihoods tabulated in Table~\ref{tab:summary of likelihoods}.
    \begin{table*}
        \centering
        \caption{List of likelihoods with different combinations of Gaussian{\iz}ation and covariance rescaling treatments, as well as different functional forms.}
        \def\arraystretch{1.2}
        \bgroup
        \setlength{\tabcolsep}{10pt}
        \begin{tabular}{rcccc}
            \toprule[0.8pt]
            \multirow{2}{*}{Data variable} & No Gaussian{\iz}ation & \multicolumn{3}{c}{With Gaussian{\iz}ation} \\
            \cmidrule[0.4pt]{2-5}
            & & \multicolumn{2}{c}{Fixed $\nu = 1/3$} & Fitted $\nu$ \\ \midrule[0.6pt]
            Functional form & Modified-$t$ & Gaussian & Modified-$t$ & Modified-$t$ \\
            \midrule[0.6pt]
            Rescaled covariance estimate (rc) & $\Like_{t,\textrm{rc}}\!\bigs{\tY}$ & $\Like_{\textrm{G}, \textrm{rc}}\!\bigs{\Z_{1/3}}$ & $\Like_{t,\textrm{rc}}\!\bigs{\Z_{1/3}}$ & $\Like_{t,\textrm{rc}}\!\bigs{\Z_\nu}$ \\
            Fixed covariance estimate (fc) & $\Like_{t,\textrm{fc}}\!\bigs{\tY}$ & -- & $\Like_{t,\textrm{fc}}\!\bigs{\Z_{1/3}}$ & $\Like_{t,\textrm{fc}}\!\bigs{\Z_\nu}$ \\
            \bottomrule[0.8pt]
        \end{tabular}
        \egroup
        \label{tab:summary of likelihoods}
    \end{table*}
All these likelihoods can be compared with the \emph{true likelihood} constructed from the exponentially distributed modes and shot noise power \emph{prior to convolution},
    \begin{equation}
        \Like_\textrm{true}(\thetap;\Y) = \prod_{i=1}^r \frac{\e^{-Y_i/P_i} - \e^{-Y_i/P_\shot}}{P_i - P_\shot} \,,
        \label{eq:true likelihood}
    \end{equation}
which is inaccessible in a realistic survey in the presence of the window function.

Since our methods mostly affect measurements on the largest survey scales, the local non-Gaussianity parameter $f_\NL$, which is sensitive to the large-scale power, is a well-motivated test parameter \citep{Sun:2013nna,Kalus:2015lna}. $f_\NL$ enters the galaxy power spectrum by modifying the constant linear galaxy bias on large scales \citep{Dalal:2007cu,Matarrese:2008nc,Slosar:2008hx},
    \begin{equation}
        b_0 \mapsto b_0 + f_\NL A(k) (b_0 - 1) \,,
    \end{equation}
which introduces scale dependence via
    \begin{equation}
        A(k) = \frac{3\varOmega_\textrm{m} \delta_\textrm{c}}{k^2 T(k)} \qty(\frac{H_0}{c})^2 \,.
    \end{equation}
Here $c$ is the speed of light, $\varOmega_\textrm{m}$ is the matter density parameter, and $\delta_\textrm{c} \approx \num{1.686}$ is the spherical collapse critical overdensity today. Henceforth we will identify $\thetap$ with $\theta = f_\NL$. We emphas{\iz}e that the choice of $f_\NL$ as a test parameter is entirely based on likelihood considerations; our work does not serve as a stringent constraint on primordial non-Gaussianity. To leading order in $f_\NL$, which is small as constrained by \textit{Planck} 2015 results \citep{Ade:2015ava}, we continue to treat galaxy overdensity as a Gaussian random field, albeit with an amplitude modulated by $f_\NL$; this will eventually break down on near the Hubble horizon scale \citep[see e.g.][]{Tellarini:2015faa}.

To properly examine the \emph{ensemble} behaviour of the likelihoods listed in Table~\ref{tab:summary of likelihoods} with different treatments, we now produce $\num{250000}$ data real{\iz}ations at some `true' input cosmology, and a fixed set of $N_\simu = \num{1000}$ mock catalogues simulated at the fiducial cosmology $f_\NL = 0$. The latter provides a covariance matrix estimate for both the band power and the Gaussian{\iz}ed data. The prior range for $f_\NL$ is set to be $[\num{-250},\num{250}]$ and scanned through with a resolution of $\Delta{f}_\NL = \num{0.05}$.\footnote{\!\!\!We have chosen a much wider prior range than the \textit{Planck} 2015 constraint (without polar{\iz}ation) $f_\NL = 2.5\pm5.7$ \citep{Ade:2015ava} to demonstrate the robustness of our likelihood treatments to a large range of power spectrum amplitudes.}

\begin{figure}
    \includegraphics[width=0.975\linewidth]{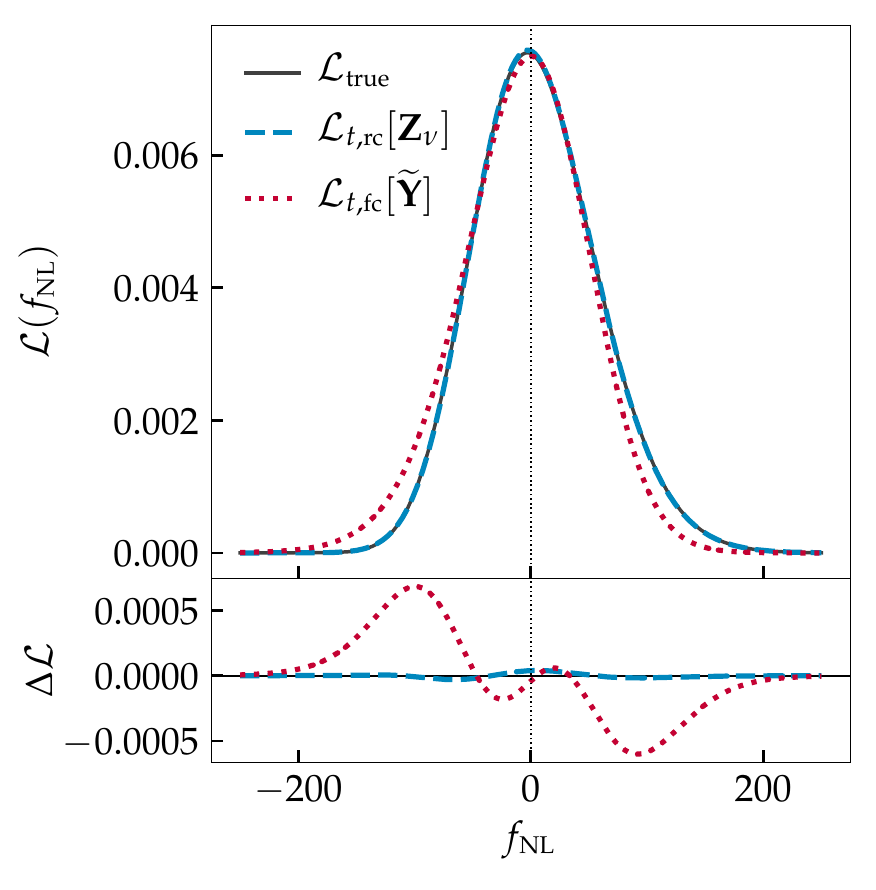}
    \caption{A direct comparison between the true likelihood $\Like_\true$ (\textit{solid line}), the modified-$t$ likelihood $\Like_{t,\textrm{rc}}\!\bigs{\Z_\nu}$ (\textit{dashed line}) \emph{with} Gaussian{\iz}ation and covariance rescaling, and the modified-$t$ likelihood $\Like_{t,\textrm{fc}}\!\bigs{\tY}$ (\textit{dotted line}) \emph{without} Gaussian{\iz}ation and covariance rescaling, all averaged over data real{\iz}ations produced at $f_\NL = 0$. The \textit{bottom panel} shows the differences of the likelihoods from $\Like_\true$.}
    \label{fig:likelihood comparison}
\end{figure}

To get an initial intuition, we compare in Fig.~\ref{fig:likelihood comparison} the true likelihood $\Like_\true$, the modified-$t$ likelihood $\Like_{t,\textrm{fc}}\!\bigs{\tY}$ without Gaussian{\iz}ation and covariance rescaling, and the new likelihood $\Like_{t,\textrm{rc}}\!\bigs{\Z_\nu}$ derived using our full procedure with fitted transformation parameter $\nu$, all averaged over data real{\iz}ations produced at $f_\NL = 0$.
It is clear that our methods produce a superior likelihood approximation to the true likelihood.

\subsubsection{Point estimation comparison}

For different true $f_\NL$ parameter inputs and the same fiducial cosmology at $f_\NL = 0$, we compare both frequentists' and Bayes estimators calculated from the likelihoods $\Like_\true$, $\Like_{t,\textrm{rc}}\!\bigs{\tY}$, $\Like_{t,\textrm{rc}}\!\bigs{\Z_\nu}$, $\Like_{t,\textrm{fc}}\!\bigs{\tY}$ and $\Like_{t,\textrm{fc}}\!\bigs{\Z_\nu}$ (see Table~\ref{tab:summary of likelihoods} for definitions). The results have been marginal{\iz}ed over our data real{\iz}ations to assess their ensemble behaviour.%
\medskip

\noindent\textit{Maximum likelihood estimator} -- This is a frequentists' estimator, given by
    \begin{equation}
        \est{\theta} = \argmax \Like(\theta) \,.
    \end{equation}
The estimates are compared in Table~\ref{tab:frequentists' estimates}. Their uncertainties are the standard deviations estimated from the ensemble of data real{\iz}ations.
\begin{table}
    \centering
    \caption{Maximum likelihood estimates of $f_\NL$ from likelihoods $\Like_\true$, $\Like_{t,\textrm{rc}}\!\bigs{\tY}$, $\Like_{t,\textrm{rc}}\!\bigs{\Z_\nu}$, $\Like_{t,\textrm{fc}}\!\bigs{\tY}$, and $\Like_{t,\textrm{fc}}\!\bigs{\Z_\nu}$ (see Table~\ref{tab:summary of likelihoods} for definitions), averaged over \num{250000} data real{\iz}ations with different true $f_\NL$ inputs and the same fiducial cosmology at $f_\NL = 0$. The 1-$\sigma$ error bounds are given as the estimated standard deviations.}
    \setlength{\tabcolsep}{4pt}
    \bgroup
    \def\arraystretch{1.1}
    \begin{tabular}{rrrrrr}
        \toprule[0.8pt]
        \multirow{2}{*}{Input} & \multicolumn{5}{c}{Maximum likelihood estimates} \\
        \cmidrule[0.4pt]{2-6}
        & \multicolumn{1}{c}{$\Like_\true$} & \multicolumn{1}{c}{$\Like_{t,\textrm{rc}}\!\bigs{\tY}$} & \multicolumn{1}{c}{$\Like_{t,\textrm{rc}}\!\bigs{\Z_\nu}$} & \multicolumn{1}{c}{$\Like_{t,\textrm{fc}}\!\bigs{\tY}$} & \multicolumn{1}{c}{$\Like_{t,\textrm{fc}}\!\bigs{\Z_\nu}$} \\
        \midrule[0.6pt]
        \num{50} & \num[separate-uncertainty]{49(42)} & \num[separate-uncertainty]{42(44)} & \num[separate-uncertainty]{48(42)} & \num[separate-uncertainty]{49(44)} & \num[separate-uncertainty]{51(42)} \\[1.25ex]
        \num{10} & \num[separate-uncertainty]{9(39)} & \num[separate-uncertainty]{2(41)} & \num[separate-uncertainty]{8(40)} & \num[separate-uncertainty]{9(39)} & \num[separate-uncertainty]{11(39)} \\[1.25ex]
        \num{0} & \num[separate-uncertainty]{-1(39)} & \num[separate-uncertainty]{-8(41)} & \num[separate-uncertainty]{-2(39)} & \num[separate-uncertainty]{-2(38)} & \num[separate-uncertainty]{1(38)}
        \\[1.25ex]
        \num{-10} & \num[separate-uncertainty]{-12(38)} & \num[separate-uncertainty]{-18(40)} & \num[separate-uncertainty]{-12(38)} & \num[separate-uncertainty]{-12(38)} & \num[separate-uncertainty]{-10(37)} \\[1.25ex]
        \num{-50} & \num[separate-uncertainty]{-52(35)} & \num[separate-uncertainty]{-58(36)} & \num[separate-uncertainty]{-52(35)} & \num[separate-uncertainty]{-52(37)} & \num[separate-uncertainty]{-50(34)} \\
        \bottomrule[0.8pt]
    \end{tabular}
    \egroup
    \label{tab:frequentists' estimates}
\end{table}%
\medskip

\noindent\textit{Posterior median estimator} -- With flat priors, the posterior $\Post[\given{\theta}{\X}]$ on the cosmological parameter(s) $\theta$ given any observations $\X$ is simply the likelihood $\Like(\theta;\X)$ suitably normal{\iz}ed. Common Monte Carlo analyses usually return the posterior median or mean as the best estimate \citep[see e.g.][]{Trotta:2008qt,Hogg:2017akh}. Here we choose the absolute loss function \citep{Berger:1985xx}
    \begin{equation}
        \textrm{loss}(a,\theta) = \abs{a - \theta}
    \end{equation}
and minim{\iz}e its expectation to obtain the Bayes estimator that is the posterior median
    \begin{equation}
        \est{\theta} = \argmin_{a} \Expc_{\given{\theta}{\X}}[\ell(a,\theta)] \,.
    \end{equation}
The associated 1-$\sigma$ uncertainties can be quoted as the equal-tailed $\SI{68.3}{\percent}$ Bayesian credible interval. The results are displayed in Table~\ref{tab:Bayes estimates}.
\begin{table}
    \centering
    \caption{Posterior median estimates of $f_\NL$ from likelihoods $\Like_\true$, $\Like_{t,\textrm{rc}}\!\bigs{\tY}$, $\Like_{t,\textrm{rc}}\!\bigs{\Z_\nu}$, $\Like_{t,\textrm{fc}}\!\bigs{\tY}$, and $\Like_{t,\textrm{fc}}\!\bigs{\Z_\nu}$ (see Table~\ref{tab:summary of likelihoods} for definitions), averaged over \num{250000} data real{\iz}ations with different true $f_\NL$ inputs and the same fiducial cosmology at $f_\NL = 0$. The 1-$\sigma$ error bounds are quoted as the equal-tailed \SI{68.3}{\percent} credible interval.}
    \setlength{\tabcolsep}{4pt}
    \bgroup
    \def\arraystretch{1.1}
    \begin{tabular}{rrrrrr}
        \toprule[0.8pt]
        \multirow{2}{*}{Input} & \multicolumn{5}{c}{Posterior median estimates} \\
        \cmidrule[0.4pt]{2-6}
        & \multicolumn{1}{c}{$\Like_\true$} & \multicolumn{1}{c}{$\Like_{t,\textrm{rc}}\!\bigs{\tY}$} & \multicolumn{1}{c}{$\Like_{t,\textrm{rc}}\!\bigs{\Z_\nu}$} & \multicolumn{1}{c}{$\Like_{t,\textrm{fc}}\!\bigs{\tY}$} & \multicolumn{1}{c}{$\Like_{t,\textrm{fc}}\!\bigs{\Z_\nu}$} \\
        \midrule[0.6pt]
        $50$ & ${53}^{+43}_{-39}$ & ${48}^{+40}_{-35}$ & ${53}^{+43}_{-38}$ & ${46}^{+33}_{-35}$ & ${51}^{+39}_{-38}$ \\[1.25ex]
        $10$ & ${13}^{+41}_{-36}$ & ${8}^{+38}_{-32}$ & ${13}^{+40}_{-35}$ & ${6}^{+36}_{-39}$ & ${11}^{+38}_{-38}$ \\[1.25ex]
        $0$ & ${3}^{+40}_{-36}$ & ${-2}^{+38}_{-32}$ & ${3}^{+40}_{-35}$ & ${-5}^{+37}_{-40}$ & ${1}^{+38}_{-37}$ \\[1.25ex]
        $-10$ & ${-7}^{+39}_{-35}$ & ${-12}^{+37}_{-31}$ & ${-7}^{+39}_{-34}$ & ${-15}^{+37}_{-41}$ & ${-9}^{+38}_{-37}$ \\[1.25ex]
        $-50$ & ${-46}^{+36}_{-31}$ & ${-52}^{+35}_{-28}$ & ${-46}^{+36}_{-30}$ & ${-56}^{+41}_{-44}$ & ${-49}^{+37}_{-37}$ \\
        \bottomrule[0.8pt]
    \end{tabular}
    \egroup
    \label{tab:Bayes estimates}
\end{table}

It is evident that overall the new likelihood $\Like_{t,\textrm{rc}}\!\bigs{\Z_\nu}$ performs the best in producing closer best estimates as well as error bounds to those from the true likelihood, whether we perform frequentists' estimation or Bayesian inference. We have found that Gaussian{\iz}ation with a fixed transformation parameter $\nu = 1/3$, i.e. $\Like_{t,\textrm{rc}}\!\bigs{\Z_{1/3}}$, gives similar results to Gaussian{\iz}ation with $\nu$ fitted by the formula in equation~\eqref*{eq:optimal transformation-parameter fitting formula}; likewise, assuming a wrong fiducial cosmology for Gaussian{\iz}ation, even when it is significantly deviant from the true cosmology, has negligible impact on recovered parameters. This demonstrates that our Gaussian{\iz}ation scheme is robust to variation of the fiducial cosmological model.

\subsubsection{Posterior shape comparison}

\begin{figure*}
    \centering
    \includegraphics[width=0.975\linewidth]{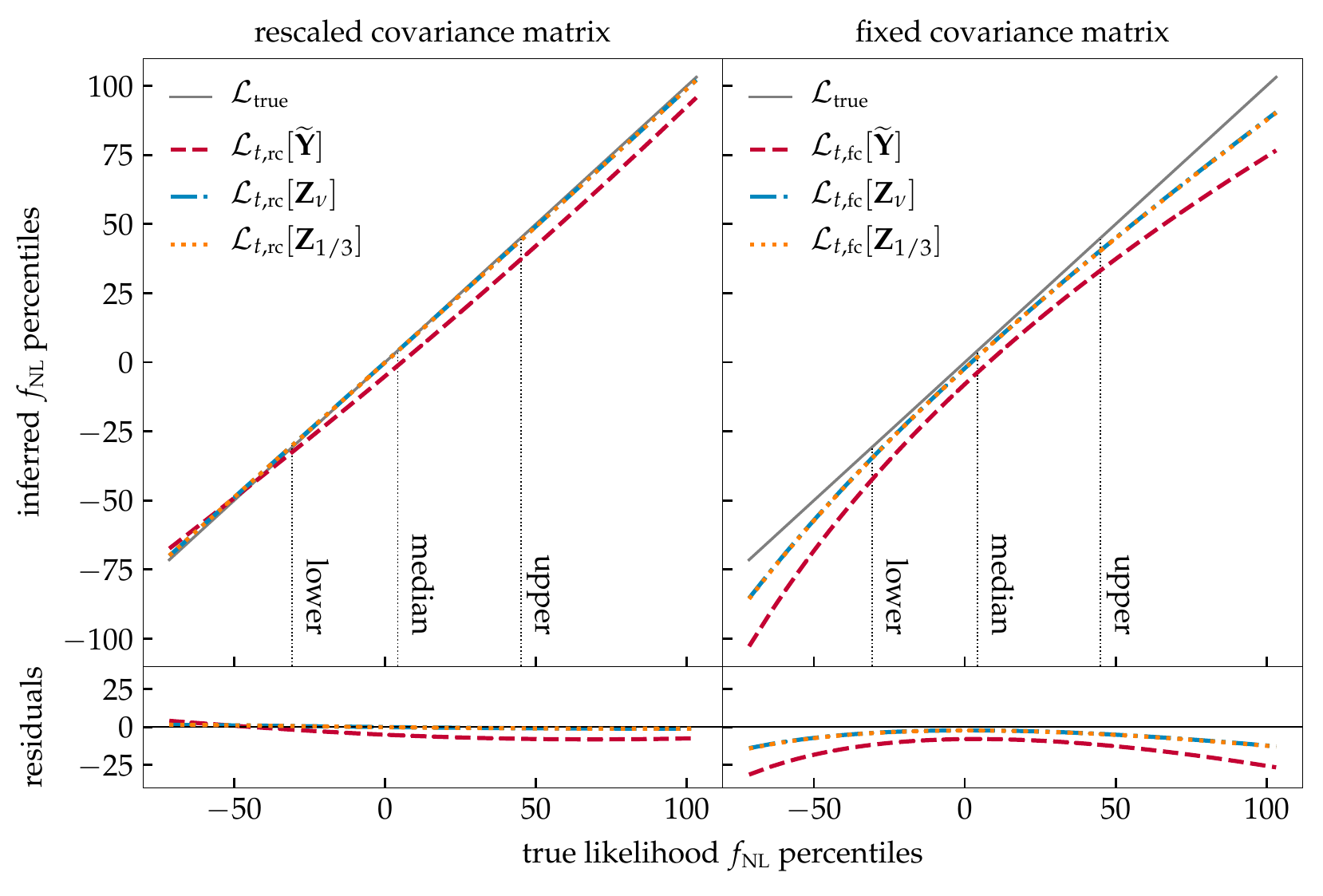}
    \caption{Q--Q plots comparing posterior distributions of the likelihoods without Gaussian{\iz}ation (data variable $\tY$, \textit{dashed line}), with fitted-$\nu$ Gaussian{\iz}ation (data variable $\Z_\nu$, \textit{dash--dotted lines}) and with fixed-$\nu$ Gaussian{\iz}ation (data variable $\Z_{1/3}$, \textit{dotted lines}) against the true posterior from $\Like_\true$ as a reference (\textit{solid lines} with unit slope). See Table~\ref{tab:summary of likelihoods} for definitions of the likelihoods. \textit{Top panel}: inferred $f_\NL$ percentiles from the different likelihoods against the true likelihood $\Like_\true$ percentiles, averaged over \num{250000} data real{\iz}ations at $f_\NL = 0$. \textit{Bottom panel}: the residuals (numerical differences) of corresponding lines from the reference line. \textit{Left-hand panel}: the fiducial covariance estimate is rescaled (`rc'). \textit{Right-hand panel}: the fiducial covariance estimate is correct but fixed in parameter space (`fc'). The dotted \textit{vertical lines} show the $f_\NL$ posterior median estimate and the 1-$\sigma$ uncertainties from the true likelihood.}
    \label{fig:probability plot}
\end{figure*}

A graphical comparison of likelihood shapes is the quantile--quantile (Q--Q) probability plot \citep{Wilk:1968} of their respective posteriors. We show the $f_\NL$ percentiles inferred from all likelihoods in Table~\ref{tab:summary of likelihoods} {except} $\Like_{\textrm{G},\textrm{rc}}\!\bigs{\Z_{1/3}}$ against $f_\NL$ percentiles of the true likelihood $\Like_\true$ in Fig.~\ref{fig:probability plot}, where we contrast no Gaussian{\iz}ation against Gaussian{\iz}ation, and fixed covariance estimates against rescaled covariance estimates.

There are two trends that match our expectation: the new likelihoods in the Gaussian{\iz}ed data variable $\Z$ matches the shape of the true likelihood $\Like_\true$ better than the ones in data variable $\tY$ without Gaussian{\iz}ation, especially away from the peak and near the tails of the distribution; not rescaling the covariance matrix in parameter space to account for its dependence on cosmology noticeably distorts the error bounds. Again we have also found that Gaussian{\iz}ation with fixed $\nu = 1/3$, i.e. $\Like_{t,\textrm{rc}}\!\bigs{\Z_{1/3}}$, produces nearly indistinguishable results from Gaussian{\iz}ation with fitted $\nu$, i.e. $\Like_{t,\textrm{rc}}\!\bigs{\Z_\nu}$.

Another quantitative measure of `statistical distance' between a true probability distribution $f(\theta)$ and an approximate probability distribution $g(\theta)$ is the Kullback--Leibler (KL) divergence \citep{Kullback:1951}
    \begin{equation}
         \KL{f}{g} \equiv \int \dd{\theta} f(\theta) \ln\frac{f(\theta)}{g(\theta)} \,.
         \label{eq:Kullback--Leibler divergence}
    \end{equation}
For instance, if we take $f$ to be the posterior of the true likelihood against which we compare the posterior $g$ of the new likelihood, then the expected KL divergence over the entire ensemble of data real{\iz}ations could quantify the `information loss' due to replacement of the true likelihood with the new.

In~Table~\ref{tab:KL divergence}, we list the KL divergence values for all likelihoods in Table~\ref{tab:summary of likelihoods}, {except} for $\Like_{\textrm{G},\textrm{rc}}\!\bigs{\Z_{1/3}}$, from the true likelihood $\Like_\true$, averaged over our data real{\iz}ations generated at different true $f_\NL$ inputs. The evidence again suggests that the likelihood $\Like_{t,\textrm{rc}}\!\bigs{\Z_\nu}$ in Gaussian{\iz}ed data with the rescaled covariance estimate matches the full shape of true likelihood very well, and this remains the case when we use Gaussian{\iz}ation at fixed $\nu = 1/3$, i.e. $\Like_{t,\textrm{rc}}\!\bigs{\Z_{1/3}}$.
\begin{table*}
    \centering
    \caption{Kullback--Leibler (KL) divergence values of $f_\NL$ posteriors of all likelihoods in Table~\ref{tab:summary of likelihoods}, \emph{except} for $\Like_{\textrm{G},\textrm{rc}}\!\bigs{\Z_{1/3}}$, from that of the true likelihood $\Like_\true$, averaged over \num{250000} data real{\iz}ations with different true $f_\NL$ inputs and the same fiducial cosmology at $f_\NL = 0$.}
    \setlength{\tabcolsep}{10pt}
    \bgroup
    \def\arraystretch{1.2}
    \begin{tabular}{rcccccc}
        \toprule[0.8pt]
        \multirow{2}{*}{Input} & \multicolumn{6}{c}{$D_\textrm{KL}$ from the true posterior} \\
        \cmidrule[0.4pt]{2-7}
        & $\Like_{t,\textrm{rc}}\!\bigs{\tY}$ & $\Like_{t,\textrm{rc}}\!\bigs{\Z_{1/3}}$ & $\Like_{t,\textrm{rc}}\!\bigs{\Z_\nu}$ & $\Like_{t,\textrm{fc}}\!\bigs{\tY}$ & $\Like_{t,\textrm{fc}}\!\bigs{\Z_{1/3}}$ & $\Like_{t,\textrm{fc}}\!\bigs{\Z_\nu}$ \\
        \midrule[0.6pt]
        \num{50} & \num{0.12} & \num{0.02} & \num{0.02} & \num{0.35} & \num{0.05} & \num{0.05} \\[1ex]
        \num{10} & \num{0.11} & \num{0.02} & \num{0.02} & \num{0.18} & \num{0.04} & \num{0.04} \\[1ex]
        \num{0} & \num{0.10} & \num{0.02} & \num{0.02} & \num{0.17} & \num{0.05} & \num{0.05} \\[1ex]
        \num{-10} & \num{0.10} & \num{0.02} & \num{0.02} & \num{0.17} & \num{0.05} & \num{0.05} \\[1ex]
        \num{-50} & \num{0.11} & \num{0.03} & \num{0.03} & \num{0.23} & \num{0.10} & \num{0.09} \\
        \bottomrule[0.8pt]
    \end{tabular}
    \egroup
    \label{tab:KL divergence}
\end{table*}

\subsection{Sources of error in parameter inference}
\label{ssec:sources of impact}

Although the major sources of impact on parameter inference, namely distribution non-normality and parameter dependence of the covariance matrix, have been identified and mitigated by Gaussian{\iz}ation and variance--correlation decomposition respectively, there are other sources of error which we now consider.

The first potential concern is how the correlations between band powers affect the Gaussian{\iz}ation. We have proposed that the Gaussian{\iz}ation be performed using only the univariate distributions, and hence this will work the best when the off-diagonal correlations in the covariance matrix are relatively weak. Thus for a given window function we want to minim{\iz}e the number of band powers to be included in the data vector so that the covariance matrix is strongly dominated by the diagonal entries. For future surveys with increasing volume, this will be easier as the window functions will be narrower in Fourier space. However, reducing the number of band powers will also mean that more overdensity modes contribute to each bin, making the statistics more Gaussian. The limit to how few band powers should be included in the data vector is that we need to make sure that there are sufficiently many bins to retain the cosmological information in the data.

The second potential concern is that when the band power variance cannot be analytically calculated from the window function mixing matrix and observed overdensity modes, the gamma distribution parameters have to be obtained from mock catalogues, and this comes with additional statistical scatter owing to the estimation of band power variance (see Section~\ref*{sssec:gamma distribution approximation}). Ideally it needs to be marginal{\iz}ed out together with the unknown full covariance matrix in the \citetalias{Sellentin:2015waz} procedure, but this unfortunately makes the likelihood analytically intractable after Gaussian{\iz}ation.

\begin{figure*}
    \centering
    \includegraphics[width=0.975\linewidth]{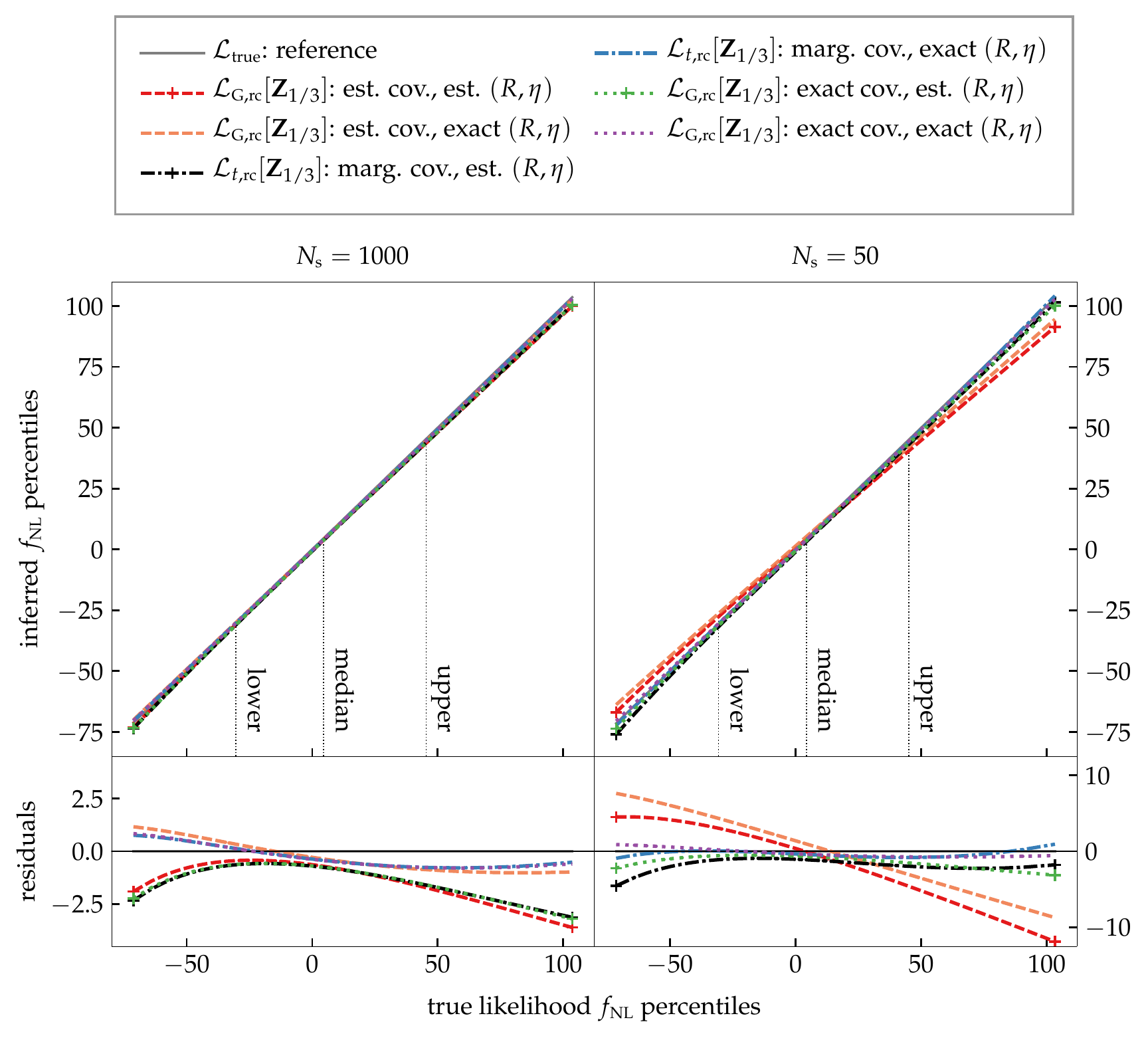}
    \caption{Q--Q plots comparing the relative impact of covariance estimation and band power variance estimation for evaluating gamma distribution parameters $(R,\eta)$. All likelihoods compared are in the Gaussian{\iz}ed data $\Z_{1/3}$ with the covariance matrix estimate rescaled for varying cosmology. The covariance matrix is either estimated without \citetalias{Sellentin:2015waz} marginal{\iz}ation (Gaussian likelihood $\Like_{\textrm{G},\textrm{rc}}\!\bigs{\Z_{1/3}}$, \textit{dashed lines}), \citetalias{Sellentin:2015waz} marginal{\iz}ed (modified-$t$ likelihood $\Like_{t,\textrm{rc}}\!\bigs{\Z_{1/3}}$, \textit{dash--dotted lines}) or exact (Gaussian likelihood $\Like_{\textrm{G},\textrm{rc}}\!\bigs{\Z_{1/3}}$, \textit{dotted lines}); the distribution parameters are either estimated (\textit{with `+' markers}) or exact (\textit{without `+' markers}). \textit{Top panel}: inferred $f_\NL$ percentiles from the different cases against percentiles of the true likelihood $\Like_\true$, averaged over \num{25000} data real{\iz}ations at $f_\NL = 0$. \textit{Bottom panel}: the residuals (numerical differences) of corresponding lines from the reference line (the true likelihood $\Like_\true$). \textit{Left-hand panel}: mock catalogues contain $N_\simu = 1000$ samples. \textit{Right-hand panel}: mock catalogues contain $N_\simu = 50$ samples. The dotted \textit{vertical lines} show the $f_\NL$ posterior median estimate and the 1-$\sigma$ uncertainties from the true likelihood. \textbf{Note} -- the scales in the bottom panels differ, and also differ from those of the bottom panels in Fig.~\ref{fig:probability plot}.}
    \label{fig:estimation impact}
\end{figure*}

To this end, we would like to assess the relative impact between covariance estimation of the Gaussian{\iz}ed band power and the estimation of the band power variance for evaluating gamma distribution parameters $(R,\eta)$. Both estimations are made from the same set of mock catalogues, so we need to consider an ensemble of mock catalogue sets. For \num{25000} data real{\iz}ations, we now generate $N_\simu = 1000$ mock catalogue samples for \emph{each} of them. In the reproduced Q--Q probability plots (Fig.~\ref{fig:estimation impact}) for likelihoods in the \emph{Gaussian{\iz}ed} data $\Z$ with fixed transformation parameter $\nu = 1/3$ and rescaled covariance estimates, we consider three scenarios of covariance estimation.
\begin{enumerate}
    \item The Hartlap-debiased precision matrix estimate is directly substituted into the Gaussian likelihood $\Like_{\textrm{G},\textrm{rc}}\!\bigs{\Z_{1/3}}$. This scenario corresponds to the dashed lines.
    \item The unbiased covariance matrix estimate is marginal{\iz}ed with the \citetalias{Sellentin:2015waz} procedure and thus the modified-$t$ likelihood $\Like_{t,\textrm{rc}}\!\bigs{\Z_{1/3}}$ is used. This scenario corresponds to the dash--dotted lines.
    \item A high-precision covariance estimate is used as a proxy for the exact covariance matrix in the Gaussian likelihood $\Like_{\textrm{G},\textrm{rc}}\!\bigs{\Z_{1/3}}$. This scenario corresponds to the dotted lines.
\end{enumerate}
The covariance estimate is rescaled for varying cosmology for each of these scenarios, and we compare using analytically calculated shape--scale parameters $(R,\eta)$ (without `+' markers) with using $(R,\eta)$ obtained from estimated band power variance (with `+' markers). In addition, we explore the effect of the mock catalogue size on the relative impacts between the two estimations, by adding the same plots (in the right-hand panel) for the case $N_\simu = 50$. The deviation from the true likelihood in these scenarios are shown as residuals (numerical differences) in the bottom panel of Fig.~\ref{fig:estimation impact}.

The evidence indicates that the \citetalias{Sellentin:2015waz} procedure (the modified-$t$ likelihood) indeed accounts for the statistical scatter of covariance estimation, but this effect is subdominant to the band power variance estimation for evaluating $(R,\eta)$ if we use a set of $N_\simu = \num{1000}$ mock catalogue samples -- we can see the lines fall into two groups, depending on whether $(R,\eta)$ are estimated or exact, in the left-hand panel of Fig.~\ref{fig:estimation impact}. If we reduce the sample size of catalogues to $N_\simu = \num{50}$, the impact of covariance estimation becomes greater than that of $(R,\eta)$ estimation -- this is evident as the lines corresponding to estimated covariance matrices deviate the most from all other lines. However, both effects are far less significant than the distribution non-normality and cosmological dependence of the covariance matrix, as we have seen in Fig.~\ref{fig:probability plot}, which are the focal problems addressed in this work. In particular, the errors due to these estimations with $N_\simu = \num{1000}$ are approaching the errors inherent in our gamma distribution approximation and univariate Gaussian{\iz}ation.

In light of these results, we recommend using the \citetalias{Sellentin:2015waz} marginal{\iz}ation procedure for the covariance matrix estimate, i.e. the modified-$t$ likelihood for parameter inference, even when we cannot do the same for the gamma distribution parameters calculated from the estimated band power variance.

\section{\texorpdfstring{A\hspace{1.10pt}p\hspace{1.10pt}p\hspace{1.10pt}l\hspace{1.10pt}i\hspace{1.10pt}c\hspace{1.10pt}a\hspace{1.10pt}t\hspace{1.10pt}i\hspace{1.10pt}o\hspace{1.10pt}n\hspace{1.10pt} P\hspace{1.10pt}i\hspace{1.10pt}p\hspace{1.10pt}e\hspace{1.10pt}l\hspace{1.10pt}i\hspace{1.10pt}n\hspace{1.10pt}e}{Application Pipeline}}
\label{sec:application}

We now present the proposed final pipeline for the straightforward application of our methods. A comprehensive list of the notations used can be found in Table~\ref{tab:notations} in \autoref*{sec:introduction}; note that in this section we use the superscript (f) to denote quantities evaluated at the fiducial cosmology, and superscript (d) for measurements or data real{\iz}ations.\medskip

\noindent\textit{Gamma distribution parameters} -- We model the band power distribution as a gamma distribution in shape--scale parametr{\iz}ation $(R, \eta)$. Given band power measurements or real{\iz}ations $\Pb^{(\data)}(k_a)$ at some cosmology $\thetap$, the shape--scale parameters are determined from its mean and variance
    \begin{equation}
        \begin{split}
            & R_a(\thetap) = \Expc\!\bigs{\Pb^{(\data)}(k_a)}^2 \Big/ \Var\!\bigs{\Pb^{(\data)}(k_a)}\,, \\
            & \eta_a(\thetap) = \Var\!\bigs{\Pb^{(\data)}(k_a)} \Big/ \Expc\!\bigs{\Pb^{(\data)}(k_a)} \,.
        \end{split}
        \label{eq:shape--scale parameters}
    \end{equation}
In the absence of analytic expressions for the band power variance, this should be replaced by a fiducial estimate $\est{\Var}\bigs{\Pb^{(\data,\fid)}(k_a)}$ calculated from mock catalogues and suitably rescaled with the cosmology $\thetap$, i.e.
    \begin{equation}
       \est{\Var}\Bigs{\Pb^{(\data)}(k_a,\thetap)} = \qty[\frac{\Pb(k_a,\thetap)}{\Pb^{(\fid)}(k_a)}]^2 \est{\Var}\Bigs{\Pb^{(\data,\fid)}(k_a)} \,,
    \end{equation}
which leads to corresponding rescaling for the distribution parameters in equation~\eqref*{eq:shape--scale parameters}. Note that this rescaling cancels out for the shape parameter $R_a$, which is in fact independent of $\thetap$. This is expected as the effective number of modes is a model-independent quantity.%
\medskip

\noindent\textit{Data transformation} -- To make the data distribution approximately multivariate normal, we adopt the Box--Cox transformation whereby the band power measurements are univariately Gaussian{\iz}ed,
    \begin{equation}
        \Pb^{(\data)}(k_a) \mapsto Z_a \equiv \Bigs{\Pb^{(\data)}(k_a)}^{\nu_a} \,.
    \end{equation}
Whilst the transformation parameters $\nu_a$ for each bin can be determined using the fitting formula given by equation~\eqref*{eq:optimal transformation-parameter fitting formula} as a function of the fiducial shape parameter $R_a^{(\fid)}$, we have found little gain over keeping this fixed at $\nu_a = 1/3$, which we favour for reasons of simplicity. After the transformation, the mean $\mu_a(\thetap)$ and variance $\sigma_a^2(\thetap)$ of the Gaussian{\iz}ed band power $Z_a$ at cosmology $\thetap$ are given by equation~\eqref*{eq:transformed mean and variance} for each bin.%
\medskip

\noindent\textit{Likelihood evaluation} -- The remaining quantity needed for likelihood evaluation is the covariance matrix estimate $\est{\mSig}(\thetap)$ for the \emph{Gaussian{\iz}ed} data $\Z$, which is allowed to vary with cosmology by rescaling the fiducial estimate $\est{\mSig}^{(\fid)}$ from $N_\simu \equiv m + 1$ mock catalogue samples,
    \begin{equation}
        \est{\mSig}(\thetap) = \mat{D} \est{\mSig}^{(\fid)} \mat{D} \,,
    \end{equation}
where the diagonal matrix $\mat{D}$ consists of entries
    \begin{equation}
        D_{aa} = \sigma_a(\thetap) \big/ \sigma_a\bigp{\thetap^{(\fid)}} \,.
    \end{equation}
We recommend using the modified-$t$ distribution obtained with \citetalias{Sellentin:2015waz} marginal{\iz}ation as the final likelihood (see equations~\ref*{eq:modified t-distribution} and \ref*{eq:full likelihood}),
    \begin{equation}
        \Like\!\bigp{\thetap;\Z,\est{\mSig}^{(\fid)},m} = \Prob_t\!\bigs{\Z;\vmu(\thetap),\est{\mSig}(\thetap),m} \,.
    \end{equation}
Our simulations in \autoref*{sec:tests} have shown that the impact from the errors in the poorly determined covariance matrix entries dominates over problems caused by the estimated gamma distribution parameters in our likelihood. Consequently, it is worth marginal{\iz}ing out scatter of the estimated covariance matrix using the \citetalias{Sellentin:2015waz} procedure even if we cannot simultaneously perform an equivalent procedure for the gamma distribution parameters.

Standard Bayesian inference can be readily performed now to extract cosmological parameter estimates and associated uncertainties, or to sample the posterior distribution in a multidimensional parameter space using Monte Carlo techniques.

\section{\texorpdfstring{C\hspace{1.10pt}o\hspace{1.10pt}n\hspace{1.10pt}c\hspace{1.10pt}l\hspace{1.10pt}u\hspace{1.10pt}s\hspace{1.10pt}i\hspace{1.10pt}o\hspace{1.10pt}n}{Conclusion}}
\label{sec:conclusion}

In preparation for next-generation galaxy surveys such as DESI and \textit{Euclid}, we have revisited the Gaussian likelihood assumption commonly found in galaxy-clustering likelihood analyses, which may adversely impact cosmological parameter inference from measurements limited by sample size on the largest survey scales. Extending previous work by \cite{Schneider:2009}, \cite{Keitel:2011}, \cite{Wilking:2013goa}, \cite{Sun:2013nna} and \cite{Kalus:2015lna}, we have carefully derived the distribution of the band power spectrum (windowed power spectrum monopole) in the linear regime while taking window effects and random shot noise into account; in particular, we have
\begin{enumerate}
    \item devised a Gaussian{\iz}ation scheme using the Box--Cox transformation to improve data normality;
    \item proposed a variance--correlation decomposition of the covariance matrix to allow for varying cosmology;
    \item presented a simple pipeline for straightforward application of this new methodology (\autoref*{sec:application}).
\end{enumerate}
We always recommend rescaling the covariance matrix estimate using our decomposition as its parameter dependence has a significant impact on parameter estimation. Although below the largest survey scales the normal distribution may be a good approximation for the band power measurements, we still recommend the use of our Gaussian{\iz}ation scheme for its simplicity.

With numerical simulations, we have tested the likelihood derived from the new procedure for both point estimation and shape comparison with the true likelihood inaccessible in real surveys. By focusing on the local non-Gaussianity $f_\NL$, which is a sensitive parameter for the large-scale power spectrum, we have demonstrated noticeable improvement in parameter inference brought by Gaussian{\iz}ation and covariance rescaling. Whilst Gaussian{\iz}ing transformations are not new, our set-up, motivation and implementation differ from previous works by, for instance, \cite{Wilking:2013goa} and \cite{Schuhmann:2015dma}.

However, an all-encompassing formalism for galaxy-clustering power spectrum analysis is still out of reach. Towards the non-linear regime where overdensity modes are no longer independent but coupled due to gravitational evolution, the power-spectrum covariance structure is fundamentally more complex, and non-negligible shot noise can also deviate from the Poisson sampling prescription \citep{Bernardeau:2001qr}. The analysis covered in this paper focuses on the windowed power spectrum monopole in the \citetalias{Feldman:1993ky} framework, but this could also be applied to power-law moment estimators with even exponents in the local plane-parallel approximation \citep{Yamamoto:2005dz}. We leave further extensions to the current analysis to future work.

\section*{Acknowledgements}

MSW, WJP, and DB acknowledge support from the European Research Council (ERC) through the Darksurvey grant 614030. SA acknowledges support from the UK Space Agency through grant ST/K00283X/1. RC is supported by the Science and Technology Facilities Council (STFC) grant ST/N000668/1.

Numerical computations are performed on the Sciama High Performance Computing (HPC) cluster which is supported by the Institute of Cosmology and Gravitation (ICG), the South East Physics Network (SEPnet) and the University of Portsmouth.

{
\raggedright
\hypersetup{urlcolor=MNRASpurple}
\bibliographystyle{mnras}
\bibliography{Bibliography}

\begin{thebibliography}{}
\makeatletter
\relax
\def\mn@urlcharsother{\let\do\@makeother \do\$\do\&\do\#\do\^\do\_\do\%\do\~}
\def\mn@doi{\begingroup\mn@urlcharsother \@ifnextchar [ {\mn@doi@}
  {\mn@doi@[]}}
\def\mn@doi@[#1]#2{\def\@tempa{#1}\ifx\@tempa\@empty \href
  {http://dx.doi.org/#2} {doi:#2}\else \href {http://dx.doi.org/#2} {#1}\fi
  \endgroup}
\def\mn@eprint#1#2{\mn@eprint@#1:#2::\@nil}
\def\mn@eprint@arXiv#1{\href {http://arxiv.org/abs/#1} {{\tt arXiv:#1}}}
\def\mn@eprint@dblp#1{\href {http://dblp.uni-trier.de/rec/bibtex/#1.xml}
  {dblp:#1}}
\def\mn@eprint@#1:#2:#3:#4\@nil{\def\@tempa {#1}\def\@tempb {#2}\def\@tempc
  {#3}\ifx \@tempc \@empty \let \@tempc \@tempb \let \@tempb \@tempa \fi \ifx
  \@tempb \@empty \def\@tempb {arXiv}\fi \@ifundefined
  {mn@eprint@\@tempb}{\@tempb:\@tempc}{\expandafter \expandafter \csname
  mn@eprint@\@tempb\endcsname \expandafter{\@tempc}}}

\bibitem[\protect\citeauthoryear{{Abbott} et~al.,}{{Abbott}
  et~al.}{2018}]{Abbott:2017wau}
{Abbott} T.~M.~C.  et~al., 2018, \mn@doi [\prd] {10.1103/PhysRevD.98.043526},
  98, 043526

\bibitem[\protect\citeauthoryear{{Ade} et~al.,}{{Ade}
  et~al.}{2016}]{Ade:2015ava}
{Ade} P.~A.~R.  et~al., 2016, \mn@doi [\aap] {10.1051/0004-6361/201525836},
  594, A17

\bibitem[\protect\citeauthoryear{{Aghamousa} et~al.,}{{Aghamousa}
  et~al.}{2016}]{Aghamousa:2016zmz}
{Aghamousa} A.  et~al., 2016, preprint
  (\href{http://arxiv.org/abs/1611.00036}{\color{blue}\texttt{arXiv:1611.00036}})

\bibitem[\protect\citeauthoryear{{Aghanim} et~al.,}{{Aghanim}
  et~al.}{2018}]{Aghanim:2018eyx}
{Aghanim} N.  et~al., {2018}, preprint
  (\href{http://arxiv.org/abs/1807.06209}{\color{blue}\texttt{arXiv:1807.06209}})

\bibitem[\protect\citeauthoryear{{Alam} et~al.,}{{Alam}
  et~al.}{2017}]{Alam:2016hwk}
{Alam} S.  et~al., 2017, \mn@doi [\mnras] {10.1093/mnras/stx721}, 470, 2617

\bibitem[\protect\citeauthoryear{Anderson}{Anderson}{2003}]{Anderson:2003xx}
Anderson T.~W.,  2003, {An Introduction to Multivariate Statistical Analysis},
  3rd edn.
Wiley

\bibitem[\protect\citeauthoryear{{Avila} et~al.,}{{Avila}
  et~al.}{2018}]{Avila:2017nyy}
{Avila} S.  et~al., 2018, \mn@doi [\mnras] {10.1093/mnras/sty1389}, 479, 94

\bibitem[\protect\citeauthoryear{Berger}{Berger}{1985}]{Berger:1985xx}
Berger J.~O.,  1985, {Statistical Decision Theory and Bayesian Analysis}.
Springer-Verlag

\bibitem[\protect\citeauthoryear{Bernardeau, Colombi, Gazta\~{n}aga  \&
  Scoccimarro}{Bernardeau et~al.}{2002}]{Bernardeau:2001qr}
Bernardeau F.,  Colombi S.,  Gazta\~{n}aga E.,   Scoccimarro R.,  2002, \mn@doi
  [\physrep] {10.1016/S0370-1573(02)00135-7}, 367, 1

\bibitem[\protect\citeauthoryear{{Beutler}, {Saito}, {Seo}, {Brinkmann},
  {Dawson}, {Eisenstein}, {Font-Ribera}  et~al.}{{Beutler}
  et~al.}{2014}]{Beutler:2013yhm}
{Beutler} F.,  {Saito} S.,  {Seo} H.-J.,  {Brinkmann} J.,  {Dawson} K.~S.,
  {Eisenstein} D.~J.,  {Font-Ribera} A.   et~al., 2014, \mn@doi [\mnras]
  {10.1093/mnras/stu1051}, 443, 065

\bibitem[\protect\citeauthoryear{{Beutler} et~al.,}{{Beutler}
  et~al.}{2017}]{Beutler:2016arn}
{Beutler} F.  et~al., 2017, \mn@doi [\mnras] {10.1093/mnras/stw3298}, 466,
  2242

\bibitem[\protect\citeauthoryear{{Blot} et~al.,}{{Blot}
  et~al.}{2019}]{Blot:2018oxk}
{Blot} L.,  et~al., 2019, \mn@doi [\mnras] {10.1093/mnras/stz507}, 485, 2806

\bibitem[\protect\citeauthoryear{Box \& Cox}{Box \& Cox}{1964}]{Box:1964xxx}
Box G.~E.~P.,  Cox D.~R.,  1964, \href{http://www.jstor.org/stable/2343588}{J.
  Royal Stat. Soc.}, 26, 211

\bibitem[\protect\citeauthoryear{Buri\'{c} \& Elezovi\'{c}}{Buri\'{c} \&
  Elezovi\'{c}}{2012}]{Burić:2011}
Buri\'{c} T.,  Elezovi\'{c} N.,  2012, \mn@doi [Integral Transforms Spec.
  Funct.] {10.1080/10652469.2011.591393}, 23, 355

\bibitem[\protect\citeauthoryear{{Colavincenzo} et~al.,}{{Colavincenzo}
  et~al.}{2019}]{Colavincenzo:2018cgf}
{Colavincenzo} M.  et~al., 2019, \mn@doi [\mnras] {10.1093/mnras/sty2964},
  482, 4883

\bibitem[\protect\citeauthoryear{D'Agostino \& Pearson}{D'Agostino \&
  Pearson}{1973}]{D'Agostino:1973}
D'Agostino R.,  Pearson E.~S.,  1973, \mn@doi [Biometrika]
  {10.1093/biomet/60.3.613}, 60, 613

\bibitem[\protect\citeauthoryear{Dalal, Dore, Huterer  \& Shirokov}{Dalal
  et~al.}{2008}]{Dalal:2007cu}
Dalal N.,  Dore O.,  Huterer D.,   Shirokov A.,  2008, \mn@doi [\prd]
  {10.1103/PhysRevD.77.123514}, 77, 123514

\bibitem[\protect\citeauthoryear{Dodelson \& Schneider}{Dodelson \&
  Schneider}{2013}]{Dodelson:2013uaa}
Dodelson S.,  Schneider M.~D.,  2013, \mn@doi [\prd]
  {10.1103/PhysRevD.88.063537}, 88, 063537

\bibitem[\protect\citeauthoryear{Duffy}{Duffy}{2014}]{Duffy:2014lva}
Duffy A.~R.,  2014, \mn@doi [Ann. Phys.] {10.1002/andp.201400059}, 526, 283

\bibitem[\protect\citeauthoryear{Eifler, Schneider  \& Hartlap}{Eifler
  et~al.}{2009}]{Eifler:2008gx}
Eifler T.,  Schneider P.,   Hartlap J.,  2009, \mn@doi [\aap]
  {10.1051/0004-6361/200811276}, 502, 721

\bibitem[\protect\citeauthoryear{Eisenstein \& Hu}{Eisenstein \&
  Hu}{1998}]{Eisenstein:1997ik}
Eisenstein D.~J.,  Hu W.,  1998, \mn@doi [\apj] {10.1086/305424}, 496, 605

\bibitem[\protect\citeauthoryear{Feldman, Kaiser  \& Peacock}{Feldman
  et~al.}{1994}]{Feldman:1993ky}
Feldman H.~A.,  Kaiser N.,   Peacock J.~A.,  1994, \mn@doi [\apj]
  {10.1086/174036}, 426, 23

\bibitem[\protect\citeauthoryear{Gabrielli, Sylos~Labini, Joyce  \&
  Pietronero}{Gabrielli et~al.}{2005}]{Gabrielli:1980xx}
Gabrielli A.,  Sylos~Labini F.,  Joyce M.,   Pietronero L.,  2005, {Statistical
  Physics for Cosmic Structures}.
Springer-Verlag

\bibitem[\protect\citeauthoryear{Golubev}{Golubev}{2016}]{Golubev:2016203}
Golubev A.,  2016, \mn@doi [J. Theor. Biol.] {10.1016/j.jtbi.2015.12.027}, 393,
  203

\bibitem[\protect\citeauthoryear{Gupta \& Nagar}{Gupta \&
  Nagar}{2000}]{Gupta:2000xx}
Gupta A.,  Nagar D.,  2000, {Matrix Variate Distributions}.
Chapman and Hall/CRC

\bibitem[\protect\citeauthoryear{Hamimeche \& Lewis}{Hamimeche \&
  Lewis}{2008}]{Hamimeche:2008ai}
Hamimeche S.,  Lewis A.,  2008, \mn@doi [\prd] {10.1103/PhysRevD.77.103013},
  77, 103013

\bibitem[\protect\citeauthoryear{Hartlap, Simon  \& Schneider}{Hartlap
  et~al.}{2007}]{Hartlap:2006kj}
Hartlap J.,  Simon P.,   Schneider P.,  2007, \mn@doi [\aap]
  {10.1051/0004-6361:20066170}, 464, 399

\bibitem[\protect\citeauthoryear{Hogg \& Foreman-Mackey}{Hogg \&
  Foreman-Mackey}{2018}]{Hogg:2017akh}
Hogg D.~W.,  Foreman-Mackey D.,  2018, \mn@doi [\apjs]
  {10.3847/1538-4365/aab76e}, 236, 11

\bibitem[\protect\citeauthoryear{Joachimi \& Taylor}{Joachimi \&
  Taylor}{2011}]{Joachimi:2011xxx}
Joachimi B.,  Taylor A.~N.,  2011, \mn@doi [\mnras]
  {10.1111/j.1365-2966.2011.19107.x}, 416, 1010

\bibitem[\protect\citeauthoryear{Kalus, Percival  \& Samushia}{Kalus
  et~al.}{2016}]{Kalus:2015lna}
Kalus B.,  Percival W.~J.,   Samushia L.,  2016, \mn@doi [\mnras]
  {10.1093/mnras/stv2307}, 455, 2573

\bibitem[\protect\citeauthoryear{Kaufman, Schervish  \& Nychka}{Kaufman
  et~al.}{2008}]{Kaufman:2008xx}
Kaufman C.~G.,  Schervish M.~J.,   Nychka D.~W.,  2008, \mn@doi [J. Am. Stat.
  Assoc.] {10.1198/016214508000000959}, 103, 1545

\bibitem[\protect\citeauthoryear{Keitel \& Schneider}{Keitel \&
  Schneider}{2011}]{Keitel:2011}
Keitel D.,  Schneider P.,  2011, \mn@doi [\aap] {10.1051/0004-6361/201117284},
  534, A76

\bibitem[\protect\citeauthoryear{{Kitaura} et~al.,}{{Kitaura}
  et~al.}{2016}]{Kitaura:2015uqa}
{Kitaura} F.-S.  et~al., 2016, \mn@doi [\mnras] {10.1093/mnras/stv2826}, 456,
  4156

\bibitem[\protect\citeauthoryear{Kullback \& Leibler}{Kullback \&
  Leibler}{1951}]{Kullback:1951}
Kullback S.,  Leibler R.~A.,  1951, \mn@doi [Ann. Math. Stat.]
  {10.1214/aoms/1177729694}, 22, 79

\bibitem[\protect\citeauthoryear{Laparra, Camps-Valls  \& Malo}{Laparra
  et~al.}{2011}]{Laparra:2011xxx}
Laparra V.,  Camps-Valls G.,   Malo J.,  2011, \mn@doi [IEEE Trans. Neural
  Netw.] {10.1109/TNN.2011.2106511}, 22, 537

\bibitem[\protect\citeauthoryear{{Laureijs} et~al.,}{{Laureijs}
  et~al.}{2011}]{Laureijs:2011}
{Laureijs} R.  et~al., 2011, preprint
  (\href{http://arxiv.org/abs/1110.3193}{\color{blue}\texttt{arXiv:1110.3193}})

\bibitem[\protect\citeauthoryear{{Li}, {Singh}, {Yu}, {Feng}  \& {Seljak}}{{Li}
  et~al.}{2019}]{Li:2018scc}
{Li} Y.,  {Singh} S.,  {Yu} B.,  {Feng} Y.,   {Seljak} U.,  2019, \mn@doi
  [\jcap] {10.1088/1475-7516/2019/01/016}, 1901, 016

\bibitem[\protect\citeauthoryear{{Lippich} et~al.,}{{Lippich}
  et~al.}{2019}]{Lippich:2018wrx}
{Lippich} M.  et~al., 2019, \mn@doi [\mnras] {10.1093/mnras/sty2757}, 482,
  1786

\bibitem[\protect\citeauthoryear{{Manera} et~al.,}{{Manera}
  et~al.}{2013}]{Manera:2013xx}
{Manera} M.  et~al., 2013, \mn@doi [\mnras] {10.1093/mnras/sts084}, 428, 1036

\bibitem[\protect\citeauthoryear{Matarrese \& Verde}{Matarrese \&
  Verde}{2008}]{Matarrese:2008nc}
Matarrese S.,  Verde L.,  2008, \mn@doi [\apjl] {10.1086/587840}, 677, L77

\bibitem[\protect\citeauthoryear{Neuts}{Neuts}{1981}]{Neuts:1981xx}
Neuts M.~F.,  1981, {Matrix-Geometric Solutions in Stochastic Models: an
  Algorthmic Approach}.
Dover Publications Inc.

\bibitem[\protect\citeauthoryear{Paz \& S\'{a}nchez}{Paz \&
  S\'{a}nchez}{2015}]{Paz:2015kwa}
Paz D.~J.,  S\'{a}nchez A.~G.,  2015, \mn@doi [\mnras] {10.1093/mnras/stv2259},
  454, 4326

\bibitem[\protect\citeauthoryear{Peacock \& Nicholson}{Peacock \&
  Nicholson}{1991}]{Peacock:1991xx}
Peacock J.~A.,  Nicholson D.,  1991, \mn@doi [\mnras]
  {10.1093/mnras/253.2.307}, 253, 307

\bibitem[\protect\citeauthoryear{Peebles}{Peebles}{1980}]{Peebles:1980xx}
Peebles P.~J.~E.,  1980, {The Large-Scale Structure of the Universe}.
Princeton University Press

\bibitem[\protect\citeauthoryear{Percival \& Brown}{Percival \&
  Brown}{2006}]{Percival:2006ss}
Percival W.~J.,  Brown M.~L.,  2006, \mn@doi [\mnras]
  {10.1111/j.1365-2966.2006.10910.x}, 372, 1104

\bibitem[\protect\citeauthoryear{{Percival} et~al.,}{{Percival}
  et~al.}{2014}]{Percival:2013sga}
{Percival} W.~J.  et~al., 2014, \mn@doi [\mnras] {10.1093/mnras/stu112}, 439,
  2531

\bibitem[\protect\citeauthoryear{Pope \& Szapudi}{Pope \&
  Szapudi}{2008}]{Pope:2007vz}
Pope A.~C.,  Szapudi I.,  2008, \mn@doi [\mnras]
  {10.1111/j.1365-2966.2008.13561.x}, 389, 766

\bibitem[\protect\citeauthoryear{Schneider \& Hartlap}{Schneider \&
  Hartlap}{2009}]{Schneider:2009}
Schneider P.,  Hartlap J.,  2009, \mn@doi [\aap] {10.1051/0004-6361/200912424},
  504, 705

\bibitem[\protect\citeauthoryear{Schuhmann, Joachimi  \& Peiris}{Schuhmann
  et~al.}{2016}]{Schuhmann:2015dma}
Schuhmann R.~L.,  Joachimi B.,   Peiris H.~V.,  2016, \mn@doi [\mnras]
  {10.1093/mnras/stw738}, 459, 1916

\bibitem[\protect\citeauthoryear{Seljak, Aslanyan, Feng  \& Modi}{Seljak
  et~al.}{2017}]{Seljak:2017rmr}
Seljak U.,  Aslanyan G.,  Feng Y.,   Modi C.,  2017, \mn@doi [\jcap]
  {10.1088/1475-7516/2017/12/009}, 1712, 009

\bibitem[\protect\citeauthoryear{Sellentin \& Heavens}{Sellentin \&
  Heavens}{2016}]{Sellentin:2015waz}
Sellentin E.,  Heavens A.~F.,  2016, \mn@doi [\mnras] {10.1093/mnrasl/slv190},
  456, L132

\bibitem[\protect\citeauthoryear{Sellentin \& Heavens}{Sellentin \&
  Heavens}{2018}]{Sellentin:2017fbg}
Sellentin E.,  Heavens A.~F.,  2018, \mn@doi [\mnras] {10.1093/mnras/stx2491},
  473, 2355

\bibitem[\protect\citeauthoryear{Sellentin, Jaffe  \& Heavens}{Sellentin
  et~al.}{2017}]{Sellentin:2017aii}
Sellentin E.,  Jaffe A.~H.,   Heavens A.~F.,  2017, preprint
  (\href{http://arxiv.org/abs/1709.03452}{\color{blue}\texttt{arXiv:1709.03452}})

\bibitem[\protect\citeauthoryear{Shao \& Zhou}{Shao \& Zhou}{2010}]{Shao:2010}
Shao Y.,  Zhou M.,  2010, \mn@doi [J. Multivar. Anal.]
  {https://doi.org/10.1016/j.jmva.2010.04.015}, 101, 2637

\bibitem[\protect\citeauthoryear{Slosar, Hirata, Seljak, Ho  \&
  Padmanabhan}{Slosar et~al.}{2008}]{Slosar:2008hx}
Slosar A.,  Hirata C.,  Seljak U.,  Ho S.,   Padmanabhan N.,  2008, \mn@doi
  [\jcap] {10.1088/1475-7516/2008/08/031}, 0808, 031

\bibitem[\protect\citeauthoryear{Smith, Challinor  \& Rocha}{Smith
  et~al.}{2006}]{Smith:2005ue}
Smith S.,  Challinor A.,   Rocha G.,  2006, \mn@doi [\prd]
  {10.1103/PhysRevD.73.023517}, 73, 023517

\bibitem[\protect\citeauthoryear{Sun, Wang  \& Zhan}{Sun
  et~al.}{2013}]{Sun:2013nna}
Sun L.,  Wang Q.,   Zhan H.,  2013, \mn@doi [\apj]
  {10.1088/0004-637X/777/1/75}, 777, 75

\bibitem[\protect\citeauthoryear{Taylor, Joachimi  \& Kitching}{Taylor
  et~al.}{2013}]{Taylor:2013xx}
Taylor A.,  Joachimi B.,   Kitching T.,  2013, \mn@doi [\mnras]
  {10.1093/mnras/stt270}, 432, 1928

\bibitem[\protect\citeauthoryear{Tegmark}{Tegmark}{1997}]{Tegmark:1996qt}
Tegmark M.,  1997, \mn@doi [\prd] {10.1103/PhysRevD.55.5895}, D55, 5895

\bibitem[\protect\citeauthoryear{Tegmark, Hamilton, Strauss, Vogeley  \&
  Szalay}{Tegmark et~al.}{1998}]{Tegmark:1997yq}
Tegmark M.,  Hamilton A.~J.~S.,  Strauss M.~A.,  Vogeley M.~S.,   Szalay A.~S.,
   1998, \mn@doi [\apj] {10.1086/305663}, 499, 555

\bibitem[\protect\citeauthoryear{Tellarini, Ross, Tasinato  \& Wands}{Tellarini
  et~al.}{2015}]{Tellarini:2015faa}
Tellarini M.,  Ross A.~J.,  Tasinato G.,   Wands D.,  2015, \mn@doi [\jcap]
  {10.1088/1475-7516/2015/07/004}, 1507, 004

\bibitem[\protect\citeauthoryear{Trotta}{Trotta}{2008}]{Trotta:2008qt}
Trotta R.,  2008, \mn@doi [Contemp. Phys.] {10.1080/00107510802066753}, 49, 71

\bibitem[\protect\citeauthoryear{{Verde} et~al.,}{{Verde}
  et~al.}{2003}]{Verde:2003ey}
{Verde} L.  et~al., 2003, \mn@doi [\apjs] {10.1086/377335}, 148, 195

\bibitem[\protect\citeauthoryear{Wilk \& Gnanadesikan}{Wilk \&
  Gnanadesikan}{1968}]{Wilk:1968}
Wilk M.~B.,  Gnanadesikan R.,  1968, \mn@doi [Biometrika] {10.2307/2334448},
  55, 1

\bibitem[\protect\citeauthoryear{Wilking \& Schneider}{Wilking \&
  Schneider}{2013}]{Wilking:2013goa}
Wilking P.,  Schneider P.,  2013, \mn@doi [\aap] {10.1051/0004-6361/201321718},
  556, A70

\bibitem[\protect\citeauthoryear{Wilson, Peacock, Taylor  \& de~la
  Torre}{Wilson et~al.}{2017}]{Wilson:2015lup}
Wilson M.~J.,  Peacock J.~A.,  Taylor A.~N.,   de~la Torre S.,  2017, \mn@doi
  [\mnras] {10.1093/mnras/stw2576}, 464, 3121

\bibitem[\protect\citeauthoryear{Wishart}{Wishart}{1928}]{Wishart:1928xx}
Wishart J.,  1928, \mn@doi [Biometrika] {10.1093/biomet/20A.1-2.32}, 20A, 32

\bibitem[\protect\citeauthoryear{Yamamoto, Nakamichi, Kamino, Bassett  \&
  Nishioka}{Yamamoto et~al.}{2006}]{Yamamoto:2005dz}
Yamamoto K.,  Nakamichi M.,  Kamino A.,  Bassett B.~A.,   Nishioka H.,  2006,
  \mn@doi [\pasj] {10.1093/pasj/58.1.93}, 58, 93

\makeatother
\end{thebibliography}
}

\appendix

\section{Shot Noise Power and Its Distribution}
\label{app:shot noise}

Here we derive the amplitude of the shot noise power and consider its distribution, which affects the power spectrum likelihood. Following the calculations in \cite{Peebles:1980xx} and \cite{Feldman:1993ky} for the two-point correlation of the Poisson-sampled overdensity field $\hat{\den}$ (see equation~\ref*{eq:realised overdensity field}), we have
    \begin{align}
        \biga{\hat{\den}(\vq) \conj{\hat{\den}}(\vq')} &= \begin{multlined}[t][0.66\linewidth]
        \int \dd[3]{\vr} \dd[3]{\vr'} \e^{\im(\vq-\vq')\vdot\vr} \e^{\im\vq'\vdot(\vr-\vr')} \\
        \times \qty[\xi(\vr-\vr') + \frac{1+\alpha}{\nbar(\vr)} \delta^\textrm{D}(\vr - \vr')]
        \end{multlined} \nonumber \\
        &= P_\true(\vq) \delta^\textrm{K}_{\vq,\vq'} + \frac{1+\alpha}{V} \!\int\! \dd[3]{\vr}\!\e^{\im(\vq-\vq')\vdot\vr} \nbar(\vr)^{-1} \,,\!\!\!
    \end{align}
where $\delta^\textrm{D}(\vr)$ is the Dirac delta function and $\delta^\textrm{K}_{\vq,\vq'}$ is the Kronecker delta function. This expectation value contains both the underlying power spectrum and the scale-invariant shot noise power
    \begin{equation}
        P_\shot = \frac{1+\alpha}{V} \int \dd[3]{\vr} \nbar(\vr)^{-1} \,.
    \end{equation}

To determine the distribution of the stochastic shot noise, we consider the scenario where $N$ galaxies are randomly located at $\{ \vx_i \}_{i=1}^N$ in a finite volume. In this set-up, the overdensity field and its Fourier transform are
    \begin{equation}
        \den(\vr) = \frac{V}{N} \sum_{i=1}^N \delta^\textrm{D}(\vr - \vx_i) - 1 \,, \quad \den(\vk) = \frac{1}{N} \sum_{i=1}^N \e^{\im\vk\vdot\vx_i}\,,
    \end{equation}
where in $\den(\vk)$ we have dropped a Dirac delta term that vanishes for $\vk \neq \vb*{0}$. In the large galaxy number limit $N \to \infty$, regardless of the detailed distribution of the summands $\exp(\im\vk\vdot\vx_i)$ where $\{ \vx_i \}_{i=1}^N$ are independently uniformly distributed, $\den(\vk)$ becomes a Gaussian random field by the central limit theorem \citep{Peacock:1991xx}. Hence the shot noise power is also exponentially distributed (cf. Section~\ref*{sssec:exact hypo-exponential distribution}), and will overlay the exponential distribution of any underlying power if there is any intrinsic structure in galaxy clustering.

\section{Hypo-exponential Distribution}
\label{app:hypo-exponential distribution}

Here we derive the form of the hypo-exponential PDF (equation~\ref*{eq:hypo-exponential distribution}) introduced in \autoref*{sec:analysis}. We will also show that the sum of independently \emph{identically} distributed exponential random variables follows the gamma distribution. This motivates the gamma distribution approximation of the hypo-exponential distribution in Section~\ref*{sssec:gamma distribution approximation}.

Let $X$ be the sum of independent exponential variables $\left\lbrace X_i\right\rbrace_{i=1}^r$ with PDF
    \begin{equation}
        \Prob_i[X_i=x_i] = \beta_i \exp(- \beta_i x_i) \,,
    \end{equation}
where $\left\lbrace \beta_i^{-1}\right\rbrace_{i=1}^r$ are their respective scale parameters. Then the PDF of $X$ is the convolution of the individual PDFs $\Prob_i$,
    \begin{multline}
        \Prob[X=x] = \qty( \Conv_{i=1}^r \Prob_i)[x] \equiv \mathop{\mathlarger{\int}} \prod_{i=1}^r \dd{t_i} \Prob_1[x-t_1] \\
        \times \prod_{j=2}^r \Prob_j[t_{j-1} - t_j] = \sum_i^r \Prob_i[x] \prod_{\substack{j \neq i \\ j = 1}}^r \frac{\beta_j}{\beta_j - \beta_i} \,.
    \end{multline}
We prove the last equality by induction on $r$: the initial statement for $r = 2$ is easy to check, so we only need to establish the inductive step
    \begin{align}
        \Prob[X=x] &= \int_0^x \dd{t} \sum_{i = 1}^{r - 1} \qty(\prod_{j \neq i}^{r - 1} \frac{\beta_j}{\beta_j - \beta_i}) \Prob_i[t] \Prob_r[x-t] \nonumber \\
        &= \e^{-\beta_r x} \sum_{i = 1}^{r - 1} \qty(\prod_{j \neq i}^{r - 1} \frac{\beta_j}{\beta_j - \beta_i}) \beta_i \beta_r \int_0^x \dd{t} \e^{-(\beta_i - \beta_r)t} \nonumber \\
        &= \sum_{i = 1}^{r - 1} \qty(\prod_{j \neq i}^{r - 1} \frac{\beta_j}{\beta_j - \beta_i}) \frac{\beta_i \Prob_r[x] - \beta_r \Prob_i[x]}{\beta_i - \beta_r} \nonumber \\
        &= \sum_{i = 1}^r \qty(\prod_{j \neq i}^r \frac{\beta_j}{\beta_j - \beta_i}) \Prob_i[x] \,.
    \end{align}

This is an example of the hypo-exponential family of distributions, sometimes also referred to as the general{\iz}ed Erlang distribution \citep{Neuts:1981xx}. We note here the particular case $\beta_i = \beta_j$ for some $i \neq j$, when two variables are also identically distributed. Using the formula above by taking the limit $\Delta\beta \equiv \beta_i - \beta_j \to 0$, the PDF of $\qty(X_i + X_j)$ is
    \begin{align}
        \Prob[X_i + X_j = x] &= \lim_{\epsilon\beta\to0} \frac{\beta_i \qty(\beta_i - \Delta\beta)}{\Delta\beta} \e^{-\beta_i x} \qty(1 - \e^{x \Delta\beta}) \nonumber \\
        &= \beta_i^2 x \e^{-\beta_i x} \,.
    \end{align}
We recogn{\iz}e this as a gamma distribution $\Gamma\bigp{2,\beta_i^{-1}}$ in the shape--scale parametr{\iz}ation.

This recovers the usual result that the sum of independently \emph{identically} distributed exponential random variables follows the gamma distribution; it also motivates our gamma distribution approximation of the hypo-exponential distribution.

\section{Error-function Transformation}
\label{app:error-function transformation}

We consider an alternative to the Box--Cox transformation adopted as our default Gaussian{\iz}ation scheme. This is derived by matching the cumulative distribution functions (CDF) and involves the (complementary) error function. Whilst this scheme is exact in principle, it requires computationally costly numerical integrations for calculating transformed moments.

To this end, we seek an invertible transformation $\wY \mapsto Z$ of the gamma random variable with fiducial shape--scale parameters $(R_\fid,\eta_\fid)$, where $Z \sim \Norm(0,1)$ is a standard normal variable with zero mean and unit variance, by matching the CDFs
    \begin{equation}
        \int_0^{\wY} \dd{t} \Prob_\Gamma[t;R_\fid,\eta_\fid] = \int_{-\infty}^Z \dd{t} \Prob_{\Norm}[t;0,1] \,,
    \end{equation}
where the gamma PDF is given by equation~\eqref*{eq:gamma distribution} and the normal PDF with zero mean and unit variance is
    \begin{equation}
        \Prob_{\Norm}[t;0,1] = \frac{1}{\sqrt{2\uppi}} \e^{- t^2/2} \,.
    \end{equation}
The solution with $\dv*{Z}{\wY} > 0$ gives the transformation
    \begin{equation}
        Z = - \sqrt{2} \Erfc^{-1}\!\qty[2 \frac{\gamma(R_\fid,\wY/\eta_\fid)}{\varGamma(R_\fid)}] \,,
        \label{eq:error-function transformation}
    \end{equation}
where $\Erfc^{-1}(x)$ is the inverse of the complementary error function
    \begin{equation}
        \Erfc(x) \equiv \frac{2}{\sqrt{\uppi}} \int_x^\infty \dd{t} \e^{-t^2} \,.
    \end{equation}

For a gamma variable $\wY$ with different shape--scale parameters $(R,\eta)$ transformed using equation~\eqref*{eq:error-function transformation}, the PDF in $Z$ is now
    \begin{equation}
        \Prob[z;R,\eta] = \Prob_{\Norm}[t;0,1] \frac{\Prob_\Gamma[\tilde{y}(z);R,\eta]}{\Prob_\Gamma[\tilde{y}(z);R_\fid,\eta_\fid]} \,,
    \end{equation}
with mean and variance given by
    \begin{equation}
        \begin{split}
            &\mu(R,\eta) = \int_{-\infty}^\infty \dd{z} z \Prob[z;R,\eta] \,, \\ &\sigma^2(R,\eta) = \int_{-\infty}^\infty \dd{z} z^2 \Prob[z;R,\eta] - \mu(R,\eta)^2 \,.
        \end{split}
        \label{eq:error-function transformed mean and variance}
    \end{equation}
These results are analogous to equation~\eqref*{eq:transformed mean and variance} for the Box--Cox Gaussian{\iz}ation scheme; however, in this case accurate evaluation of transformed moments requires computationally expensive numerical integration for each parameter pair $(R,\eta)$. For this reason we do not implement this scheme in our pipeline.

\section{Covariance Marginalization}
\label{app:covariance marginalisation}

We now show that the covariance matrix decomposition, which is used in rescaling the fiducial covariance estimate to allow for cosmological dependence, does not affect the \citetalias{Sellentin:2015waz} procedure for marginal{\iz}ing out the scatter due to covariance estimation using simulated data samples.

Let us consider the unbiased estimator of the true covariance matrix $\mSig \in \R^{p \times p}$ for $(m + 1)$ samples $\left\lbrace\X_i\right\rbrace_{i=1}^{m+1}$ of a random vector $\X \in \R^p$,
    \begin{align}
        \est{\mSig} &= \frac{1}{m+1} \sum_i \bigp{\X_i - \overbar{\X}} \trans{\bigp{\X_i - \overbar{\X}}} \,, \nonumber \\
        \hspace{-2ex}\qq{where} \overbar{\X} &= \frac{1}{m+1} \sum_i \X_i \,.
    \end{align}
The distribution of $\est{\mSig}$ conditional on $\mSig$ is Wishart $\Wishart_p(\mSig/m,m)$ with the PDF \citep{Wishart:1928xx}
    \begin{align}
        \Prob_{\Wishart}\!\bigs{\bgiven{\est{\mSig}}{\mSig}} = \frac{\abs{\mSig/m}^{-m/2}}{2^{mp/2} \varGamma_p(m/2)} &\big\vert{\est{\mSig}}\big\vert^{(m-p-1)/2} \nonumber \\
        \times &\exp[-\frac{m}{2} \tr(\mSig^{-1} \est{\mSig})] \,,
    \end{align}
where the multivariate gamma function $\varGamma_p$ is defined by
    \begin{equation}
        \varGamma_p(x) \equiv \uppi^{(p-1)p/4} \prod_{j=1}^p \varGamma\qty(x + \frac{1-j}{2}) \,.
    \end{equation}
We also introduce the inverse Wishart distribution $\Wishart_p^{-1}\!\bigp{m\est{\mSig},m}$ with the PDF
    \begin{align}
        \Prob_{\WishartInv}\!\bigs{\bgiven{\mSig}{\est{\mSig}}} = \frac{\big\vert{m\est{\mSig}}\big\vert^{m/2}}{2^{mp/2} \varGamma_p(m/2)} &\abs{\mSig}^{-(m+p+1)/2} \nonumber \\
        \times &\exp[-\frac{m}{2} \tr(\est{\mSig} \mSig^{-1})] \,.
    \end{align}
Both the Wishart and inverse Wishart distribution possess the following invariance property \citep{Gupta:2000xx}: given a non-singular matrix $\mat{D} \in \R^{p \times p}$,
\begin{enumerate}
    \item if $\boldsymbol{\mathsf{\Psi}} \sim \Wishart_p(\mat{A}, m)$, then $\trans{\mat{D}} \boldsymbol{\mathsf{\Psi}} \mat{D} \sim \Wishart_p(\trans{\mat{D}} \mat{A} \mat{D}, m)$;
    \item if $\boldsymbol{\mathsf{\Psi}} \sim \Wishart^{-1}_p(\mat{A}^{-1}, m)$, then $\trans{\mat{D}} \boldsymbol{\mathsf{\Psi}} \mat{D} \sim \Wishart_p(\trans{\mat{D}} \mat{A}^{-1} \mat{D}, m)$.
\end{enumerate}

Since in our covariance matrix decomposition the rescaling matrix is the diagonal matrix of standard deviations (see equations~\ref*{eq:covariance decomposition} and \ref*{eq:covariance calibration}), the cosmology-varying covariance matrix $\mSig(\thetap)$ has the inverse Wishart posterior distribution $\Wishart_p^{-1}\!\bigp{m\est{\mSig}(\thetap),m}$,
    \begin{align}
        \Post\!\bigp{\bgiven{\mSig(\thetap)}{\est{\mSig}(\thetap)}} = \frac{\big\vert{m\est{\mSig}}\big\vert^{m/2}}{2^{mp/2} \Gamma_p(m/2)} &\abs{\mSig}^{-(m+p+1)/2} \nonumber \\
        \times &\exp[-\frac{m}{2} \tr(\est{\mSig}\mSig^{-1})] \,.
    \end{align}
This shows that the variance--correlation decomposition does not change the \citetalias{Sellentin:2015waz} marginal{\iz}ation step in Section~\ref*{ssec:covariance treatment}.

Finally, by marginal{\iz}ing the normal distribution over the posterior distribution of $\bgiven{\mSig}{\est{\mSig}}$ derived above, we can replace the unknown covariance matrix $\mSig(\thetap)$ with an unbiased estimate $\est{\mSig}(\thetap)$ from $(m+1)$ data samples. This leads to the modified $t$-distribution (equation~\ref*{eq:modified t-distribution}) introduced in \cite{Sellentin:2015waz}, which we recommend using as the likelihood form in our analysis pipeline.

\bsp

\end{document}